\documentclass[11pt]{article}
 \usepackage[a4paper,hmarginratio=1:1,vmarginratio=2:3,totalwidth=15.6cm,
  totalheight=21.4cm]{geometry}

\usepackage{amsmath}
\usepackage{amssymb}
\usepackage{amsfonts}
\usepackage{epsfig}

\newcommand{\cL}{{\cal L}}
\newcommand{\cM}{{\cal M}}

\newcommand{\cO}{{\cal O}}

\newcommand{\cW}{{\cal W}}

\newcommand{\ra}{\rightarrow}
\newcommand{\be}{\begin{equation}}
\newcommand{\ee}{\end{equation}}
\newcommand{\bea}{\begin{eqnarray}}
\newcommand{\eea}{\end{eqnarray}}

\DeclareMathSymbol{\mg}{\mathrel}{symbols}{"1D}

\newcounter{oldcounter}
\addtocounter{equation}{1}
\begin{document}
 \begin{flushright}
{CPHT-RR084.0807, LPT-Orsay 07-59}\\
{CERN-PH-TH/2007-133, OUTP-0707P}
\end{flushright}
\thispagestyle{empty}
\vspace{0.3cm}
\begin{center}
{\Large {\bf Supersymmetric Models with Higher Dimensional\\
\bigskip
 Operators}}\vspace{1.cm}

 {\bf I. Antoniadis$^{\,a, b}$, E. Dudas$^{b, c}$,
  D.~M. Ghilencea$^{\,d}$}\\
\vspace{0.5cm}
 {\it $^a $Department of Physics, CERN - Theory Division, 1211 Geneva 23,
 Switzerland.}\\[6pt]
 {\it $^b $CPHT, UMR du CNRS 7644,  \'Ecole Politechnique, 91128
   Palaiseau Cedex, France.}\\[6pt]
 {\it $^c $LPT, UMR du CNRS 8627, B\^at 210, Universit\'e de Paris-Sud,
 91405 Orsay Cedex, France}
 \\[6pt]
 {\it $^d $Rudolf Peierls Centre for Theoretical Physics, University of
   Oxford,\\
 1 Keble Road, Oxford OX1 3NP, United Kingdom.}\\

 \end{center}
 \begin{abstract}
\noindent 
In 4D renormalisable theories, integrating out massive states
generates in the low energy effective action higher dimensional operators
(derivative or otherwise).
Using a superfield language it is shown that a  4D N=1
supersymmetric theory  with  higher derivative operators  in
either the Kahler or the superpotential part of the Lagrangian and
with an otherwise arbitrary superpotential, is equivalent to  a 4D
N=1 theory of second order (i.e. without higher derivatives) with
additional superfields and renormalised interactions.
We provide examples where a free theory with  trivial supersymmetry
 breaking provided by a linear superpotential becomes,  in the
 presence of higher derivatives terms and in the second order version, 
a non-trivial interactive one with spontaneous supersymmetry
breaking. 
The couplings of the equivalent theory acquire 
a threshold correction through their dependence on the scale of 
 the higher dimensional operator(s).
The scalar potential in the second order theory is not necessarily positive
definite, and one can in principle have a vanishing potential with 
broken supersymmetry.  We provide an application to MSSM
 and argue that at tree-level and for a mass scale associated to
a higher derivative term  in the TeV range, the Higgs mass can be 
lifted above the current experimental limits.
\end{abstract}

\newpage
\setcounter{page}{1}

\section{Introduction}

The search for Physics beyond the Standard Model (SM) within
the framework of effective field theories addresses
in particular the role  of  higher dimensional
operators and their possible experimental footprints.
In  effective field theories and in models compactified
to 4D these operators are a common presence. A special class of
operators is that of higher derivative operators, whose
role has not been investigated in great detail. In this work
both classes of higher dimensional  operators (derivative or
otherwise) are included. One  motivation for
considering the  study of such operators is
that they  can be generated at low energy (below some scale)
 by renormalisable new physics at this scale,  
after integrating out massive degrees of freedom. 
Such operators are also generated dynamically  by radiative corrections 
even in the simplest orbifold compactifications, see for example
\cite{nibbelink,Ghilencea:2004sq,Ghilencea:2006qm,santamaria,Alvarez:2006we}.
Such operators can  be  generated by either bulk
or brane-localised interactions, when integrating out
the loop corrections of modes associated with the compactification.
 For example they are generated by one-loop gauge  interactions
 in 6D compactifications
\cite{nibbelink,Ghilencea:2006qm,santamaria,Alvarez:2006we}
and by one-loop localised superpotential interactions  in 5D or
6D orbifolds \cite{Ghilencea:2004sq}.
These operators respect all the symmetries of the models considered
and their generation as counterterms by quantum effects prompted,
in part,  the present analysis.

In the context of effective field theories higher dimensional
derivative operators   were investigated in the past; they were
studied in the framework  of   Randall-Sundrum models \cite{RS},
have implications for  cosmology
\cite{Armendariz-Picon:1999rj,Anisimov:2005ne,Gibbons:2003yj},
phase transitions and Higgs models \cite{BMP,c1,c5,GGP},
 supergravity/higher derivative gravity
\cite{Stelle}-\cite{Avramidi:1985ki}, string theory
\cite{Eliezer:1989cr,Polyakov:1986cs}, cosmological constant
\cite{Kaplan:2005rr}, implications for the UV regime
\cite{Antoniadis:2006pc}, for  instabilities \cite{Nesterenko:2006tu},
 and model building \cite{Grinstein:2007mp}-\cite{Buchmuller:2004eg}.
 Applications of theories
with such operators also included their role in
regularisation methods  debated in
\cite{Slavnov:1977zf,Martin:1994cg,Asorey:1995tq,Bakeyev:1996is}.
Interacting theories with higher derivatives involve the presence
of  ghosts which can bring in difficult issues (for example
unitarity violation), some of which were
studied in \cite{Smilga:2005gb,ArkaniHamed:2003uy}. Such issues are  actually 
common, since the presence of ghosts is also familiar
in standard Pauli-Villars regularisation method of 4D theories
(see \cite{Leon:1995nm} and references therein).
Provided that the scale of the higher derivative operators
is high-enough the associated effects are suppressed at low energies.
In theories with higher derivative
operators the vacuum-to-vacuum amplitude is well-defined
(no exponential growth) provided that the ghost fields are 
not asymptotic states \cite{Hawking} (i.e. are present as loop states
 only).  In this case the vacuum-to-vacuum amplitude (and therefore 
Green functions) tends to that of second order theory in the decoupling
limit of a very high  scale of these operators. One looses
unitarity near this high scale, but can never produce a negative-norm 
state or negative probability \cite{Hawking}.

The purpose of this work is to study
 the case of generic 4D N=1 supersymmetric models
 with higher derivative terms  in either its Kahler
 part or in the superpotential, for an otherwise arbitrary
superpotential and field content.
Using a supersymmetric  formulation it is shown that such models
are in fact  equivalent to a second order theory
with  renormalised interactions and additional
(ghost) superfields manifestly present in the action.
The couplings of the new,
second order theory, acquire dynamically, at the tree level,
a  dependence  on the scale of the higher
dimensional operators, via wavefunction rescaling.
For specific assumptions for the analytical continuation of the
theory from Minkowski to Euclidean metric and in the
absence of additional higher dimensional non-derivative operators, we
argue that the new theory can be renormalisable for the case of
higher derivative terms considered.

Supersymmetry breaking is also considered and in this higher
derivative operators can play an interesting role. It is showed that 
apparently un-interesting models with higher derivative terms,
without interactions and with trivial supersymmetry breaking in
 the decoupling limit of the higher derivative terms, are in 
fact interacting in the new, second order formulation,
and also have spontaneous supersymmetry breaking 
\`a la  O'Raifeartaigh \cite{o'r}.

Independent of the exact nature of the ghost (super)fields mentioned
earlier  (loop only or asymptotic states) the  method we  develop enables us to 
estimate perturbatively  the effects of high-scale physics due to
higher  derivative operators on low energy observables.
The  presence of  ghost superfields
leads to a scalar potential which is not necessarily positive
definite and one could in principle have a positive, negative
or even  vanishing  scalar potential for broken  supersymmetry.
The last case can be relevant for the  cosmological constant problem \cite{Weinberg}.

We would like to emphasize one point, briefly mentioned in the first paragraph,
which we consider extremely  important regarding
the origin of the higher dimensional operators (derivative or
otherwise). 
Consider 
a 4D  renormalisable theory with a massive superfield $\chi$,
of mass $M_*\gg m$, with
\bea
\cL_1=\int d^4\theta\, \Big[\Phi^\dagger \Phi +\chi^\dagger \chi\Big]
+\bigg\{
\int d^2\theta \bigg[\frac{M_*}{2}\,\chi^2+m\,\Phi\,\chi+
\cW(\Phi)\bigg]+h.c.\bigg\}
\eea
After solving the eq of motion for $\chi$, one immediately finds
\medskip
\bea
\chi
&=&
\frac{1}{M_*}\Big[-m\,\Phi-\frac{m}{4M_*}\,{\overline D}^2\Phi^\dagger
+\frac{1}{16}\,\frac{-m}{M^2_*}\,{\overline D}^2 \,D^2\Phi
-\frac{m}{64\, M^3_*}\,\overline D^2\, D^2\,\overline D^2\Phi^\dagger
+\cdots\Big]\quad
\eea
which if plugged back into the Lagrangian brings about terms of the 
form
\medskip
\bea\label{www}
\cL_1&\supset &
\int d^4\theta\,\bigg\{
\frac{m^2}{8\,M_*^3} \,\Big[\Phi\,D^2\,\Phi+h.c.\Big]
+\frac{m^2}{16\, M_*^4}\,(\overline D^2\Phi^\dagger)\, (D^2 \Phi)+\cdots
\bigg\}
\eea
Therefore, integrating out massive (super)fields generated higher derivative
operators in the low energy {\it effective} action valid below the
scale $M_*$. Since 
the original theory was free of ghosts, the same remains  true
about the low energy action {\it as long as} one considers the whole
{\it series} of higher dimensional operators in (\ref{www}). 
However, in an effective theory
study  and for practical purposes,  one is often 
restricting the analysis to a finite set of higher dimensional 
operators, of lowest order in $1/M_*$.
 The consequence of this is the presence of ghosts in
the action, as a sign that  the theory is incomplete in the UV
(i.e. physics above $M_*$).
From an effective field theory point of view, as we
adopt in this paper, the
absence of a UV completion is assumed anyway, therefore  we are 
not addressing, in our discussion below, 
the  more conceptual problems that ghosts 
can eventually bring. Our goal is
to show however,  how  one can investigate {\it effective} theories with 
higher derivative operators in difficult 
 cases such as when the original, high energy  action is unknown.
For a  discussion of related issues see \cite{ADGT}.

The plan of the paper is as follows. Section~\ref{sec0} introduces the
lowest-order higher derivative terms
in superfield language.  In Section~\ref{sec1} we
consider a 4D N=1 model with higher derivative kinetic term and show
its equivalence to a second order theory. The analysis is done
firstly for a specific superpotential, to set out our method of
``unfolding'' the fourth order theory into an equivalent, second
order one. This  is then generalised to an arbitrary
superpotential (without derivative terms). In Section~\ref{sec3} we
discuss the case of higher derivative terms in the superpotential
which is otherwise arbitrary, and perform a similar analysis, to
find the equivalent second order theory. The case of spontaneous
supersymmetry breaking and the form of soft terms
is briefly discussed in Section~\ref{sec4}. 
In Section \ref{sec6} we discuss briefly an application to
 MSSM with higher derivative operators  and show that the lightest Higgs
 mass acquires corrections of order $2\mu/M_*$,  which are sizable 
for mass scales $M_*$  suppressing the operator in the 10 TeV range and 
that can raise the Higgs mass above the current experimental bounds.
We end with the conclusions and a short Appendix.

\section{Higher dimensional operators~: general considerations}\label{sec0}

The framework of our analysis is that of 4D N=1 supersymmetric models.
In such models,  one can commonly have
higher dimensional operators, which can involve  higher derivatives
or not. Let us consider the last case.
 Despite rendering a model non-renormalisable,
such operators are important for phenomenology.
Our analysis in this paper is valid in the presence of
such operators in the superpotential and does not make explicit
reference to them. If such operators are present in the
Kahler term, they again do not affect the analysis below.
This is because  our analysis  will only
involve transformations (change of ``basis'') of fields having
higher  (super)derivatives and of their derivatives,
to a new basis of superfields where such derivatives are absent.
Since we will provide the relations
between the two bases of fields, old and new, the higher dimensional
operators referred  to earlier, if they involve a field
which\footnote{having a (super)derivative} underwent such a
transformation,  will immediately be
known in the new, equivalent basis. This will become clearer in the
next sections. Briefly, higher dimensional operators are
 spectators under the transformations we consider, allowing
our analysis to be  general.

Let us now consider, in the 4D N=1 supersymmetric context,
the possible operators involving (super)derivatives, to the lowest orders.
Such operators are less studied in the literature.
They can be constructed using  combinations of powers of
$D^2, \overline D^2$ and  of the superfields denoted $\Phi_j$.
For the lowest powers of
$D$, $\overline D$, these are
(integrated appropriately over Grassmann space):
\medskip
\bea\label{derivs}
&&(a)\,\int d^2\theta \,\Phi_i\,(\overline D^2\,\Phi_j^\dagger)+h.c.,
\qquad
(b)\, \int d^4\theta\,[\Phi_i\,(D^2\Phi_j)+h.c.]\sim \int
d^2\theta\,\Phi_i\,\Box \,\Phi_j+h.c\nonumber\\[10pt]
&&(c)\,\int d^4\theta\,\Phi^\dagger_i\,\overline D^2\,D^2\,\Phi_j
\sim \int d^4\theta\,\Phi^\dagger_i\,\Box \,\Phi_j,\qquad
(d)\,\int d^4 \theta\,\Phi_i^{\dagger} \ \Phi_j D^2 \Phi_k
\qquad
\eea

\medskip
\noindent (a) is just $\int d^4\theta(\Phi_i^\dagger \Phi_j+h.c.)$, (b) has
dimension 5, (c) and (d) have dimension 6; (b) is studied in
section~\ref{sec3};
(c) is studied in section~\ref{sec1} while (d) can be treated in a
similar way, see section~\ref{sec4}.
Further, one could also consider
\bea
(e)\,\int d^2\theta\,\Phi^n\,(\overline
D^2\Phi_j^\dagger)^p+h.c.,
\eea
of dimension $2p+n+1$.
This  generalises  (a) and can also be treated following
the same method as in Section~\ref{sec3} and for $n+p\leq 3$ can 
actually be renormalisable. In fact
(b), (c), (d), can also be renormalisable (see Section~\ref{ren})
despite having dimension $5$, $6$, $6$ respectively. The
renormalisability will be seen once we write a theory with
such operators as an equivalent, second order theory involving only
dimension 4 operators. The renormalisable  character of the new,
equivalent theory,  adds support to their study and partly motivated
their analysis.

\section{Effects of higher derivative kinetic terms.}\label{sec1}

\subsection{The Wess-Zumino model with higher derivative terms.}
\label{WZnohdo}

We start with the Wess-Zumino Lagrangian with a higher derivative term,
with  $\Phi\equiv(\phi,\psi,F)$:
\medskip
\begin{eqnarray}\label{WZaction}
\cL&=&\int d^4\theta \,\Phi^\dagger \Big(1+ \xi \,\Box\Big)\Phi
+\bigg\{\int d^2\theta \bigg[\frac{1}{2} \,m \Phi^2 + \frac{1}{3}\lambda
  \,\Phi^3\bigg]+ c.c.\bigg\}
\nonumber\\[9pt]
&=&
F^* \Big( 1+\xi {\Box} \Big) \,F
- \phi^*\Box \Big( 1+\xi {\Box} \Big) \,\phi
+i \partial_\mu \bar \psi  \,\bar\sigma^\mu
\Big( 1+\xi {\Box} \Big)  \psi\nonumber\\[9pt]
&+&\Big( \,
\frac12  m \,\big( 2 \phi\,F -\psi\psi \big)+\lambda\,
\big(\phi^2 \,F-\phi\,\psi^2\big)
+c.c.\Big)
\end{eqnarray}

\medskip
\noindent
where $\xi\equiv 1/M_*^2$, with $M_*$ the scale where the higher
dimensional operator becomes relevant. We assume that $M_*^2$ is
significantly
larger than all other scales in the theory (like
$m^2$ or vev's of the fields).
Due to the presence of the
 term proportional to $\xi$, the auxiliary field  $F$ is now {\it
   dynamical}\footnote{By supersymmetry, this will require, in a second
   order formulation of this theory to be found in the following, the
   introduction of an additional superfield, see later}.
The spectrum of scalar states is found from the poles of the
propagators $\langle\phi\,\phi^*\rangle$, $\langle F \,F^*\rangle$.
From these, one finds the  masses as the roots of
%\bea
$\Box\,(1+\xi\,\Box)^2+m^2=0,\,\,$
%\eea
given by
\medskip
\bea\label{spc}
m_1^2 &=& m^2\,\Big[1+2\,\xi\,m^2 +
7\,(\xi\,m^2)^2+\cO\Big((\xi\,m^2)^{5/2}\Big)\Big]\nonumber\\[8pt]
m_{2,3}^2 &=&
m^2\,\bigg[\frac{1}{\xi\,m^2}\pm \frac{1}{\sqrt{\xi\,m^2}}
-\frac{1}{2}\pm \frac{5\sqrt{\xi\,m^2}}{8}-\xi\,m^2+\cO\Big((\xi\,m^2)^{3/2}\Big)\bigg]
\eea

\medskip
\noindent
The last two masses correspond to two ghost states
 associated with $F$ and $\Box\phi$, see their negative kinetic
terms. These values will be needed in the
 next section. For later reference,  the
 scalar potential in the limit $\xi=0$ is:
$V(\phi)=\vert m\,\phi+\lambda\,\phi^2\vert^2$
which has two supersymmetric ground states $V=0$, $\langle
F\rangle =0$ situated at $\langle\phi\rangle=\langle\phi^*\rangle=0$ and $\langle
\phi\rangle=-m/\lambda$,  $\langle\phi^*\rangle=-m/\lambda^*$. The mass matrix in the basis
$(\phi,\phi^*)$  has eigenvalues $m_{1,2}^2=m^2$. A saddle point
is located at $\phi=-m/(2\,\lambda)$, $\phi^*=-m/(2\,\lambda^*)$, where  $m_{1,2}^2=\pm m^2/2$.
One can use the above Lagrangian for calculations, including loop
effects  in the presence of the higher derivative operator
\cite{Antoniadis:2006pc}. However,
it would be preferable to have a better understanding of the role of
such operator and, if possible,
 a formulation of such models as a second
order theory. This could prove  very helpful for  applications.

\subsection{The higher derivative Wess-Zumino model as a second order theory}
\label{hdosec}

In this section  we show in a manifest supersymmetric way
 that a Wess-Zumino model with a higher derivative term is
 equivalent to  a second order theory with new
superfields and renormalised
interactions. The results are then generalised to an
 arbitrary superpotential.  One has (here\footnote{Only $s_1=+1$ gives
 a bounded Euclidean action, but  we keep $s_1$ only to trace its
 effects in formulae below.}
 $s_1\equiv\pm 1$)
\medskip
\begin{eqnarray}\label{originalL}
\cL&=&\int d^4 \theta\, \Phi^\dagger \Big(1+s_1 \,\xi\,\Box\Big) \Phi +
\bigg\{ \int d^2 \theta \,\bigg[\frac{m}{2}\, \Phi^2 +
\frac{\lambda}{3}\, \Phi^3\bigg] +h.c.\bigg\}\nonumber\\[12pt]
&=&
\int d^4 \theta\, \Big[
\Phi^\dagger \Phi - \frac{s_1 \,\xi}{16} \,\,D^2 \Phi\,\overline D^2
\Phi^\dagger
\Big]
+\bigg\{\int d^2 \theta \,\bigg[\frac{m}{2}\, \Phi^2 +
\frac{\lambda}{3}\, \Phi^3\bigg] +h.c.\bigg\}
\end{eqnarray}

\medskip
\noindent
Introduce
\begin{eqnarray}\label{trasf}
\Phi &=& a_1\, \Phi_1+a_2\, \Phi_2\nonumber\\
\Phi_D\equiv m^{-1}\, \overline D^2\, \Phi^\dagger &=& b_1\, \Phi_1+b_2\, \Phi_2,
\end{eqnarray}

\medskip
\noindent
where we used that $\Phi_D\equiv (\phi_D,\psi_D,F_D)$
 is  itself a chiral superfield.
 The $2\times 2$ matrix of coefficients $a_{1,2}$, $b_{1,2}$ must be
 unitary, to maintain the eigenvalue
 problem.  A useful parametrisation  
for  the unitary matrix is
 $a_1=\cos\theta\,\exp(i\,h_1)$, $a_2=\sin\theta \exp
(-i(h-h_1))$, $b_1=-\sin\theta\,\exp(i\,(h-h_1))$,
$b_2=\cos\theta\,\exp(-i\,h_1)$ where  $\theta,\,h,\,h_1$ are 
real. 
In principle one could work with a simplified assumption ($a_1=b_2=1$,
$a_2=b_1=0$) since the difference is a rotation in the superfield
space. For generality we keep the matrix entries in the above non-trivial
parametrization, to show explicitly that the final results are 
independent of  a such particular choice
\footnote{For an easier, more transparent first reading one  could set in the 
following  $a_1=b_2=1$, $a_2=b_1=0$.}.
 Eq.(\ref{trasf})  gives a constraint
\medskip
\begin{eqnarray}\label{constraint}
 m^{-1} \bigg[a_1^* {\overline D}^2\,\Phi_1^\dagger +a_2^*
{\overline
    D}^2\,\Phi_2^\dagger
\bigg]=b_1 \,\Phi_1+b_2 \,\Phi_2
\end{eqnarray}

\medskip
\noindent
To account for this, we must  introduce an additional
contribution  $\Delta L$ to the Lagrangian,
where the Lagrange multiplier is a new chiral
superfield $\Phi_3$ (therefore $\overline D \Phi_3=0$). We then 
have\footnote{Without any restriction of generality, we used
the scale $m$ in eq.(\ref{trasf}), (\ref{constraint}), 
introduced for dimensional reasons. In principle one can use
there any other finite, {\it non-zero} mass scale of the theory,
$m_q$.
However, our use of $m$ instead of $m_q$
 only amounts to a simple  re-definition, always allowed (and assumed
 to have been made), of our original parameter $\xi\ra \xi \,m_q^2/m^2$,
as it can easily be seen  from inserting (\ref{trasf}) in 
 eq.(\ref{originalL}).}

\medskip
\begin{eqnarray}\label{delta-L}
\Delta \cL=
\int d^2\theta\,
\,\Big[m^{-1}\,
(a_1^* {\overline D}^2\,\Phi_1^\dagger +a_2^* {\overline
    D}^2\,\Phi_2^\dagger)
- ( b_1 \,\Phi_1+b_2 \,\Phi_2 )\Big]\, \Phi_3\,m_* +h.c.,\quad
m_*\equiv \frac{\sqrt{\xi} \,m^2}{4}
\end{eqnarray}

\medskip
\noindent
Above $m_*$ was introduced for dimensional reasons; since the constraint
should be removed in the absence of the higher derivative term $(\xi\ra 0$)  in the
original action, $m_*$ should be proportional to $\xi$; further, each
of the $D^2\Phi$ or $\overline D^2\Phi^\dagger$
derivatives comes with a $\sqrt{\xi}/4$, see eq.(\ref{originalL}),
and with these observations one then obtains  the above expression  for $m_*$.
 With $\cL'=\cL+\Delta \cL$, then
\medskip
\begin{eqnarray}\label{LL}
\cL'&=&
\int d^4 \theta\,\,\Big[
\rho_1 \,\Phi_1^\dagger \Phi_1 +
\big(\rho_2 \, \Phi_1^\dagger \Phi_2 +\rho_2^* \Phi_2^\dagger \Phi_1 \big)
+ \big( \rho_3\, \Phi_1^\dagger \,\Phi_3+
\rho_4 \,\Phi_2^\dagger \,\Phi_3+h.c. \big)
+\rho_5 \,\Phi_2^\dagger \Phi_2
\Big]\nonumber\\[12pt]
&+&\!\!\!\!
\bigg\{
\int d^2 \theta \,\Big[\frac{m}{2}\, \big(a_1\Phi_1+a_2\Phi_2 \big)^2 +
\frac{\lambda}{3}\, \big(a_1 \Phi_1+a_2\Phi_2 \big)^3
-m_* \Phi_3 \, \big(b_1\, \Phi_1+b_2\,\Phi_2 \big)
\Big] +h.c.\bigg\}\qquad
\end{eqnarray}

\medskip
\noindent
where
\begin{eqnarray}\label{cis}
\rho_1& = & \vert a_1\vert^2
-\frac{s_1}{16}\,\vert b_1\vert^2\,\xi\,m^2,\qquad
\rho_2 = a_1^*\,a_2
-\frac{s_1}{16}\,b_1^*\,b_2\,\xi\,m^2, \nonumber\\[12pt]
\rho_3 & = & - 4\,\frac{m_*}{m}\,a_1^*, \qquad\qquad\qquad\quad
\rho_4= -4\,\frac{m_*}{m}\,a_2^*,\qquad \qquad
\rho_5  =  \vert a_2\vert^2
 -\frac{s_1}{16}\,\vert b_2\vert^2\,\xi\,m^2
\end{eqnarray}

\medskip
\noindent
We therefore ``traded'' the
 higher derivative term in the original action,
for an additional superfield $\Phi_2$ plus a constraint, which
generated in turn the presence of $\Phi_3$.
After using  the eq of motion (in terms of superfields) for
$\Phi_3\equiv(\phi_3,\psi_3,F_3)$ one immediately 
finds\footnote{with 
$-4\int d^4 x \, d^4 \theta\, f(x,\theta,\overline\theta)=\int d^4 x\,
 d^2\theta \,\overline D^2\,f(x,\theta,\overline\theta)$, $f$ 
arbitrary.}$^,$\footnote{For 
later use let us mention that in our conventions
\begin{eqnarray}\label{sr}
\overline D^2 \Phi^\dagger \equiv \big(-4 F^*; \, -4 i \, \partial
\!\!\! \slash \overline \psi;\, 4 \,\Box \phi^*\big), \qquad 
\Phi\equiv (\phi,\psi,F)
\end{eqnarray}} (after using the definition of $\rho_{3,4}$) that
\smallskip
\begin{eqnarray}\label{con1}
m_*\bigg\{ m^{-1} \Big[a_1^* {\overline D}^2\,\Phi_1^\dagger +a_2^*
{\overline
    D}^2\,\Phi_2^\dagger
\Big]-\Big[b_1 \,\Phi_1+b_2 \,\Phi_2\Big]\bigg\}=0.
\end{eqnarray}

\medskip
\noindent
For $m_*\not\!=\!0$, this immediately recovers  the initial constraint
(\ref{constraint}), while if
 $m_*\propto\sqrt\xi\ra 0$  the constraint
$\Delta\cL$ of (\ref{delta-L}) is vanishing, as it should be the case
since in this case there is no higher derivative term in
(\ref{originalL}).

Also, from eq.(\ref{LL}) one can integrate out $\Phi_2$, $\Phi_3$ by using
the their equation of motion,
to  recover the original lagrangian (\ref{originalL}).
To see this easily one  uses their eqs of motion in superfield form and
that: $-4\int d^4 x\,  d^4 \theta\, f(x,\theta,\overline\theta)=\int d^4 x
 \, d^2\theta \,\overline D^2\,f(x,\theta,\overline\theta)$ 
and $\overline D^2 D^2=-16\Box $.

We diagonalise the hermitian matrix $\cM$ of the coefficients of
  D terms in (\ref{LL}), in the basis\footnote{$\cM$ has:
$\cM_{11}\!=\!\rho_1, \cM_{12}\!=\!\rho_2, \cM_{13}\!=\!\rho_3,
\cM_{21}\!=\!\rho_2^*,
 \cM_{22}\!=\!\rho_5,  \cM_{23}\!=\!\rho_4, \cM_{31}\!=\!\rho_3^*,
 \cM_{32}\!=\!\rho_4^*, \cM_{33}\!=\!0$} $(\Phi_1,\Phi_2,\Phi_3)$.
Its eigenvalues $\nu_{1,2,3}$ are  real, given by the roots of
\bea\label{roots1}
\nu^3+\nu^2\,c_3+\nu \,c_2+c_1=0;
\eea
where
\begin{eqnarray}
c_3&=&-\frac{1}{16}\Big[16 \Big( \vert a_1\vert^2+\vert
  a_2\vert^2\Big)
- s_1 \,\Big(\vert b_1\vert^2+\vert b_2\vert^2
\Big)\,\xi\,m^2 \Big];\nonumber\\[9pt]
c_2 &=&-
\frac{1}{16}
\Big[256 \,\frac{m_*^2}{m^2}   
\, \Big(\vert a_1\vert^2+\vert
  a_2\vert^2\Big)
+ s_1 \,\,\vert a_2 b_1- a_1 b_2 \vert^2\,\, \xi \,m^2
\Big]
 \nonumber\\[9pt]
c_1 &=& -s_1 \,\frac{m_*^2}{m^2}\,\,\vert a_2 \,b_1-a_1\,b_2\vert^2
\,(\xi\, m^2)
\end{eqnarray}
These expressions show explicitly the invariance of $c_i$ under
unitary redefinitions of the $2\times 2$ matrix of coefficients 
$a_i$,$b_i$, and this is a good consistency check.
If $s_1>0$,  there is one positive root
and two negative.
If $s_1\!<\!0$ we end up with two positive
and one negative roots.
For a  unitary transformation in (\ref{trasf}), the roots are
\medskip
\begin{eqnarray}\label{roots2}
\nu_{1,2}=\frac{1}{2}\Big[1\pm\sqrt{1+64 \,m_*^2/m^2}\Big],\,\,\textrm{with}\,\,\,
\nu_1>0;\qquad
\nu_3=-\frac{1}{16}\,s_1 \, m^2\, \xi;
\end{eqnarray}

\medskip
\noindent
where as usual $m_*=\sqrt{\xi}\,m^2/4$.
We keep $m_*$ manifest in our
equations in order to trace the effects of the initial constraint,
eqs.(\ref{constraint}) and (\ref{delta-L}).
The Lagrangian becomes:
\medskip
\begin{eqnarray}\label{LL2}
\cL'&=&
\int d^4 \theta\,\,
\bigg[
\nu_1\,\Phi_1^{' \dagger} \Phi_1^{'}
+
\nu_2\,\Phi_2^{' \dagger} \Phi_2^{'}
+
\nu_3\, \Phi_3^{' \dagger} \Phi_3^{'}
\bigg]
\nonumber\\[12pt]
&+&
\bigg\{
\int d^2 \theta \,\bigg[\frac{1}{2}\, m_{kp} \,\Phi_k' \,\Phi_p'
+
\frac{1}{3}\,\lambda_{kpl}\,\Phi_k'\,\Phi_p'\,\Phi_l'
\bigg] +h.c.\bigg\}
\end{eqnarray}

\medskip
\noindent
where  $\Phi_i'=v_{ij} \,\Phi_j$, $i,j=1,2,3$,
 and  ${\rm diag}(\nu_1,\nu_2,\nu_3)=v\,\cM\,v^\dagger$ with $v_{ij}$ unitary. Also
\medskip
\bea\label{newmasses}
m_{kp}& = & m\,\Big( a_1 \,v_{k1}^* +a_2 \,v_{k2}^* \Big)
\Big( a_1 \,v_{p1}^* +a_2 \,v_{p2}^* \Big)
- \,m_*\,\Big( v_{k3}^* (b_1 \,v_{p1}^*+
b_2 \,v_{p2}^* )+  (k\leftrightarrow p)\Big)
\nonumber\\[10pt]
\lambda_{kpl}& =& \lambda \,
\Big( a_1 \,v_{k1}^* +a_2 \,v_{k2}^* \Big)
\Big( a_1 \,v_{p1}^* +a_2 \,v_{p2}^* \Big)
\Big( a_1 \,v_{l1}^* +a_2 \,v_{l2}^* \Big)\label{ww3}
\eea

\medskip
\noindent
which are symmetric under the permutation of their indices.
We  rescale $\tilde\Phi_i'$
\begin{eqnarray}\label{rescale1}
\Phi'_i=
\tilde\Phi_i/\sqrt{\vert \nu_i\vert},\qquad i=1,2,3.
\end{eqnarray}
to find:
\medskip
\begin{eqnarray}\label{LL3}
\cL'&=&
\int d^4 \theta\,\,
\bigg[
\sigma_{\nu_1}\,\tilde\Phi_1^{\dagger} \tilde\Phi_1
+
\sigma_{\nu_2}\,\tilde\Phi_2^{\dagger} \tilde \Phi_2
+
\sigma_{\nu_3}\, \tilde\Phi_3^{\dagger}\tilde\Phi_3
\bigg]
\nonumber\\[12pt]
&+&
\bigg\{
\int d^2 \theta \,\bigg[\frac{1}{2}\, \tilde m_{kp} \,\tilde\Phi_k \,
\tilde\Phi_p
+
\frac{1}{3}\,\tilde\lambda_{kpl}\,\tilde\Phi_k\,\tilde\Phi_p\,\tilde\Phi_l
\bigg] +h.c.\bigg\},
\eea
where
\bea
\sigma_{\nu_k}=\frac{\nu_k}{\vert\,\nu_k\,\vert},\qquad
\tilde m_{kp}=\frac{m_{kp}}{\sqrt{\vert \nu_k \,\nu_p\,\vert}},
\qquad
\tilde\lambda_{kpl}=\frac{\lambda_{kpl}}{\sqrt{\vert \nu_k
    \,\nu_p\,\nu_l\vert}},
\qquad k,p,l=1,2,3.
\eea

\medskip
\noindent
Therefore $\sigma_{\nu_1}=1$, $\sigma_{\nu_2}=-1$,
$\sigma_{\nu_3}=-s_1$ and as a result $\tilde\Phi_1$ is a
particle-like field, $\tilde\Phi_2$ and $\tilde\Phi_3$ (for $s_1=1$)
are ghost-like superfields given their negative kinetic terms.
The presence of such superfields is  common  in
supersymmetric theories with constraints \cite{Siegel}.

The result in eq.(\ref{LL3})  shows that we  ``unfolded'' in a
manifest  supersymmetric way the original, higher derivative
supersymmetric Lagrangian  $\cL$ eqs.(\ref{originalL}) into an
equivalent, second order Lagrangian.
As a result, while in the initial (\ref{originalL})
the auxiliary field $F$ was {\it dynamical},  all new $\tilde F_i$ in $\cL'$
 are not and can be integrated out as usual.
To understand this change, recall from (\ref{sr}) the components of
$\Phi_D \equiv (\phi_D,\psi_D, F_D)\!\sim\! (-4 F^*; \, -4 i \,
\partial \!\!\! \slash \overline \psi;\, 4 \,\Box \phi^*)$; further,
original $(\Phi, \Phi_D)$ were traded for $(\Phi_1,\Phi_2)\ra
(\tilde\Phi_1, \tilde\Phi_2)$;  using these components of $\Phi_D$
we see then that  $\tilde\Phi^\dagger_2\tilde\Phi_2$ in (\ref{LL3})
accounts for the kinetic term of original  $F$ and for the  higher
derivative terms for original $\psi, \phi$ in (\ref{originalL}).
However, since each of the components of 
$\Phi_D$ were not independent of those of  $\Phi$, a constraint had
to  account for this, which was ``traded'' for a new chiral
superfield $\tilde\Phi_3$. In this case $\tilde\Phi_3$ is dynamical,
but we
  shall see in Section 4 that this is not always the 
case\footnote{It is perhaps useful to mention here
 the  non-susy situation, discussed long ago in \cite{Hawking}.
In this case taking $L=- 1/2 \phi (\Box + \xi \Box^2) \phi$, and
after  introducing the lagrangian multiplier $\lambda$ one
 finds $L'=-1/2\, \phi
  \rho-1/2 \,\xi\,\rho^2+\lambda (\Box \phi-\rho)$ thus the 
field $\rho$ is not dynamical and can be integrated out
(alternatively one can factorise the ``$\Box$''
 dependence in $L$ to end up with two dynamical 
fields \cite{Hawking}). Therefore only two fields are present in the end,
$\phi$ and $\lambda$.  This  is different 
from the supersymmetric case in the text where original $F$ was dynamical,
which by supersymmetry required the introduction of an additional
 (third) (super)field in the second order formulation.}.
Further
\bea
 \sigma_{\nu_j}\, \tilde F_j^{*}=-
\Big[\tilde m_{kj} \,\tilde\phi_k+ \tilde\lambda_{kpj}
\,\tilde\phi_k\,\tilde\phi_p\Big],\qquad j=1,2,3.
\eea

\medskip\noindent
The scalar potential of the ``unfolded'' theory is\footnote{Sums over
 repeating indices are understood.}:
\smallskip
\bea\label{v0}
V      &=& \sigma_{\nu_j}\, \vert \tilde F_j\vert^2
=\,
\sigma_{\nu_j}
\,
\vert
\tilde m_{kj} \,\tilde\phi_k+ \tilde\lambda_{kpj}
\,\tilde\phi_k\,\tilde\phi_p \,\vert^2
\qquad \,k,p,j=1,2,3.\qquad
\eea

\medskip\noindent
Therefore the scalar potential V is {\it not} positive
definite  in  the second order theory; it has contributions
 which are negative due to  ghost superfields.

\subsection{The mass spectrum.}

Let us investigate the mass spectrum.
One obtains for the trilinear couplings (\ref{newmasses})
\medskip
\bea\label{newcoupling}
\tilde \lambda_{111}=\tilde
\lambda_{221}=-\tilde\lambda_{112}=-\tilde\lambda_{222}=
-\frac{\lambda}{ {\eta^{3/4}}},
\qquad \tilde\lambda_{ij3}=0, \qquad
i,j=1,2,3,
\eea
symmetric under a permutation of their indices; also
\bea\label{eta1}
\eta\equiv 1+64\,m_*^2/m^2
\eea
For the bilinear couplings
\bea\label{newmass}
\tilde m_{11}&=&\tilde m_{22}=
 -\tilde m_{12}=\frac{m}{\sqrt\eta},\qquad\qquad
\tilde m_{33}=0 \nonumber\\[9pt]
 \tilde m_{13}&=&\frac{1-\sqrt{\eta}}{
2\sqrt{\xi} \,\,\eta^{1/4}},
\qquad\qquad\qquad \qquad  % \nonumber\\[9pt]
\tilde m_{23}=- \frac{1+\sqrt\eta}{2\sqrt{\xi}\,\,
\eta^{1/4}},
\eea

\medskip\noindent
 Note
 that\footnote{For $m_{13}$ and $m_{23}$ one also obtains an
overall phase factor
$\exp(-i\,h_1)\,sign(\sec\theta)$ multiplying the values shown
in (\ref{newmass}),  and which was not written there since it can be
absorbed into a redefinition of $\tilde\Phi_3$, see (\ref{LL5}).}
$m_{23}\sim -1/\sqrt{\xi}$ while the rest are
finite when $\xi\ra 0$  (recalling that $m_*=m^2\sqrt{\xi}/4$).
With
the above relations, the Lagrangian simplifies into
\medskip
\begin{eqnarray}\label{LL5}
\cL'&=&
\int d^4 \theta\,\,
\bigg[
\tilde\Phi_1^{\dagger} \tilde\Phi_1
-\tilde\Phi_2^{\dagger} \tilde \Phi_2
-s_1
\tilde\Phi_3^{\dagger}\tilde\Phi_3
\bigg]
\\[12pt]
&+&
\bigg\{
\int d^2 \theta \,\bigg[
(\tilde m_{13}\,\tilde\Phi_1+\tilde m_{23}\,\tilde\Phi_2)\,\tilde\Phi_3
+
\frac{\tilde m_{11}}{2}
  \,(\tilde\Phi_1-\tilde\Phi_2)^2+
\frac{\tilde\lambda_{111}}{3}\,(\tilde\Phi_1- \tilde\Phi_2)^3
\bigg] +h.c.\bigg\},\nonumber
\eea

\medskip
\noindent
which shows that $\tilde\Phi_3$ has no trilinear interaction.

The tree-level couplings  ($\tilde m_{ij}$ and
$\tilde\lambda_{ijk}$) of the new theory have acquired
a scale (moduli) dependence.
 Here we refer to their dependence on $\xi$ which
is explicit  in eqs.(\ref{newmass}) and to that induced via
$m_*\sim \sqrt \xi$, where $m_*$
is the parameter controlling the presence of
the constraint $\Delta \cL$, eqs.(\ref{delta-L}), (\ref{eta1}).
Therefore the constraint itself introduced dynamically a  scale
dependence of the couplings. This dependence is ultimately
due to the  higher derivative operators (\ref{originalL})
whose initial presence in the Lagrangian was ``traded'' for a
threshold correction to the couplings of the Lagrangian of
the second order  theory.

The relation of initial fields
$\Phi, \,\Phi_D$ to the new basis is,
for the unitary transformation (\ref{trasf})
\medskip
\bea\label{eee}
\Phi=\frac{1}{\eta^{1/4}}\,(\tilde\Phi_2-\tilde\Phi_1),\qquad\quad
\Phi_D=\frac{4}{m\,\sqrt{\xi}} \,\,\tilde\Phi_3
\eea

\medskip
\noindent
This shows that the  original superfield $\Phi$ has actually
a ``ghost-like'' component ($\tilde\Phi_2$); note that the overall factor
$\eta$ depends  on the scale of the higher derivative operator.

From (\ref{LL5}) one finds  the scalar
potential:
\medskip
\bea\label{pot4}
V& =& \big\vert \tilde F_1\big\vert^2 -
\big\vert \tilde F_2\big\vert^2 - s_1\,
\big\vert \tilde F_3\big\vert^2
\nonumber\\[9pt]
&=&
\big(\tilde m_{13}^2-\tilde m_{23}^2\big)\,\big \vert\, \tilde\phi_3\,\big \vert^2
+
 \,\big(\tilde m_{13}+\tilde m_{23}\big)\,\Big\{
\big[\tilde \phi_3^* \,\big(\tilde
 m_{11} \,\tilde \phi_-  +\tilde \lambda_{111} \,\,\tilde
 \phi_-^2\,\big)\big]
+c.c.\Big\}
\nonumber\\[9pt]
&-& s_1\,
\big\vert \,\tilde m_{13}\,\tilde
 \phi_1 +
 \tilde m_{23}\,\tilde \phi_2\,\big\vert^2
\eea

\medskip\noindent
 The quartic interaction is not present anymore in the potential of the
``unfolded'' Lagrangian, which only contains bilinear and cubic terms.
Tree-level quartic interactions are nevertheless generated in the low
 energy limit by exchange of $\tilde\phi_3$.
 This form  of the potential seems unstable
due to cubic terms present and would suggest that  such stability
be only  addressed  locally (i.e. we demand that  $\xi\,\tilde\phi_i^2\ll 1$)
and   other higher dimensional operators of similar
 order could  affect the Lagrangian at large $\tilde\phi_i$.
In fact, the discussion of stability in ghost directions is rather subtle as we
 shall see later, due to the fact that
these fields have negative kinetic terms.
We return to clarify this  shortly.
The vanishing of the first derivatives
 wrt $\tilde\phi^*_{1,2,3}$ respectively, gives
\medskip
\bea\label{re1}
\quad && \langle\tilde\phi_3\rangle =\frac{-1}{\tilde m_{13}-\tilde m_{23}}
 \,\bigg[\tilde m_{11} +\tilde \lambda_{111}
   \Big(\langle \tilde\phi_1\rangle-\langle\tilde\phi_2\rangle\Big)\bigg]\,
\Big(\langle\tilde\phi_1\rangle-\langle\tilde\phi_2\rangle\Big)
\\[14pt]
&&\tilde m_{13}\,\langle\tilde \phi_1 \rangle+\tilde m_{23}
\,\langle\tilde \phi_2\rangle=0,\quad
\textrm{and}\quad
\label{re2}
\\[12pt]
i) &\!\!&\!\! \langle \tilde \phi_1\rangle-\langle\tilde\phi_2\rangle=0,
\quad\textrm{or}\quad
 ii) \,\, \,\langle\tilde \phi_1\rangle-\langle\tilde\phi_2\rangle=-\frac{\tilde
  m_{11}}{\tilde\lambda_{111}},\quad\textrm{or}\quad
iii)\,\,\, \langle\tilde \phi_1\rangle-
\langle\tilde\phi_2\rangle =-\frac{\tilde
  m_{11}}{2\tilde\lambda_{111}}\label{re3}\quad
\eea

\medskip\noindent
These cases are discussed below.
In case $i)$ we have, using (\ref{re1}), (\ref{re2}):
\bea\label{casei}
\langle\tilde F_{1,2,3}\rangle=0,\qquad\langle\tilde \phi_{1,2,3}\rangle=0,\qquad   V_*=0
\eea
where $V_*$ denotes the value of the scalar potential at this
extremum point.
One computes the eigenvalues of the mass matrix of second
derivatives of the scalar potential  $V$, which is expanded about this
vacuum solution, in the basis $(\tilde\phi_i,\tilde\phi_i^*)$, $i=1,2,3$.
At this extremum point one finds (for $s_1=+1$)
\smallskip
\bea\label{w1}
m^2_{\tilde{\tilde \phi}_1}&=&
m^2\Big[\, 1-7\,(\xi\,m^2)^2+ \cO\Big((\xi\,m^2)^{5/2}\Big)\Big]\nonumber\\[10pt]
m^2_{\tilde {\tilde\phi}_{2,3}}&=&m^2\Big[\,
-\frac{1}{\xi\,m^2}\mp \frac{1}{\sqrt{\xi} m}-\frac{1}{2} \pm \frac{19
  \sqrt{\xi}\, m}{8}+\cO({\xi\,m^2})\Big]
\eea

\medskip
\noindent
where either the upper or lower signs are to be considered,
for $\tilde\phi_{2,3}$ respectively.
The first eigenvalue should correspond to our original
 $\phi$ in (\ref{WZaction}),
(\ref{originalL}) of mass $m$, for the same supersymmetric
 state. There are also two
negative  mass eigenstates  corresponding to the two ghost superfields
 present in the ``unfolded'' Lagrangian of second order.
Their negative signs are expected since the {\it kinetic and mass
  terms}  of ghost superfields
come with opposite sign in the action and thus do not necessarily suggest an
instability of the potential in the vicinity of this vacuum.
There is however a  problem. The above spectrum is  different from
that in (\ref{spc}) of the original Lagrangian (\ref{WZaction}),
although the latter is equivalent to that in (\ref{LL5}) if
the two formulations
are indeed equivalent, as showed. What is, then,
the origin of this discrepancy?

To understand this,  note that -  unlike above -
one should  compute the mass eigenvalues from
the  potential with a  metric
which takes account of the different  sign
of the ghosts' kinetic terms. For this one goes to the basis
$\phi_i=(a_i+i b_i)/\sqrt 2$, $\phi_i^*=(a_i-i b_i)/\sqrt 2$
 where $a_i, b_i$  $\,(i=1,2,3)$ are real components,
then rescale $a_i, b_i$ for $i=2,3$,
 into $a_i\ra i \,a_i$, $b_i\ra i \,b_i$, ($a_1, b_1$ fixed).
This rescaling   ensures positive definite kinetic terms for the ghost terms.
In the basis $(a_1, b_1, i a_2, i b_2, i a_3, i b_3)$
 any negative eigenvalue of the mass matrix will
signal a local instability.
In this new basis the matrix of second derivatives of the
potential has eigenvalues controlled by the characteristic equation
\bea \label{sstt}
 \Big(\Box \,(1+\xi\,\Box)^2+m^2\Big)=0
\eea
which is identical to that  discussed in the text after
eq.(\ref{WZaction}). Therefore, the spectrum of the original Lagrangian
(\ref{WZaction}), (\ref{originalL}) of the theory with higher
derivative operators
is indeed  identical to that of the second order theory,
computed after an appropriate  rescaling of their {\it real} components,
to account for their initial negative kinetic terms.
 This is a good
check of the equivalence of the two formulations of the
theory, eqs.(\ref{WZaction}) and (\ref{LL5}),  and of the introduction of the
additional constraint superfield in (\ref{delta-L}). It also shows
explicitly the {\it requirement}  for using a different
field-space metric when computing
the spectrum  in a second order theory with ghost fields.
These observations fix the problem mentioned above and invalidate
 (\ref{w1}).  The correct spectrum is then
\medskip
\bea\label{w1p}
m^2_{\tilde{\tilde \phi}_1} &=& m^2\,\Big[1+2\,\xi\,m^2 +
7\,(\xi\,m^2)^2+\cO\Big((\xi\,m^2)^{5/2}\Big)\Big]\nonumber\\[8pt]
m^2_{\tilde {\tilde\phi}_{2,3}} &=&
m^2\,\bigg[\frac{1}{\xi\,m^2}\pm \frac{1}{\sqrt{\xi\,m^2}}
-\frac{1}{2}\pm \frac{5\sqrt{\xi\,m^2}}{8}-\xi\,m^2+\cO\Big((\xi\,m^2)^{3/2}\Big)\bigg]
\eea

\medskip
\noindent
in  agreement with (\ref{spc}).  In the basis of rescaled
component fields of positive definite kinetic terms
all mass eigenvalues are positive and this vacuum is therefore stable.

In case $ii),\,$  using (\ref{re1}), (\ref{re2}):
\medskip
\bea\label{caseii}
\langle\tilde F_{1,2,3}\rangle=0,
\qquad\langle\tilde \phi_{1}\rangle
=\frac{m\,(1+\sqrt\eta)}{2\eta^{1/4} \lambda},\qquad
\langle\tilde \phi_{2}\rangle=\frac{m\,(1-\sqrt\eta)}{2\eta^{1/4}\lambda},
\qquad \langle\tilde\phi_3\rangle=0,\qquad V_*=0
\eea

\medskip
\noindent
with the vev of $\tilde\phi_2$ going to 0 if $\xi\,m^2\ra 0$.
In this case
the eigenvalues of the matrix of second derivatives of the scalar
potential
are identical to those  of case i),  eq.(\ref{w1p}).
The same situation as in $i)$ applies regarding the stability.
Supersymmetry is unbroken in both $i)$ and $ii)$, and in the
limit $\xi\ra 0$ one recovers from $i), ii)$ the two
supersymmetric ground states of
the Wess-Zumino models without higher dimensional operators.

Finally, consider case $iii)$  together with
 (\ref{re1}), (\ref{re2}). One obtains
\medskip
\bea\label{rtq}
\langle\tilde F_1^*\rangle
=-\frac{ m^2 (1+\sqrt{\eta})}{8\lambda \eta^{1/4}},\qquad
\langle\tilde F_2^*\rangle=-\frac{ m^2 (1-\sqrt{\eta})}{8\lambda \eta^{1/4}},\qquad
\langle\tilde F_3\rangle=0
\eea
and
\bea
\langle\tilde \phi_{1}\rangle=\frac{m\,(1+\sqrt\eta)}{4\eta^{1/4}\lambda},\qquad
\langle\tilde \phi_{2}\rangle=\frac{m\,(1-\sqrt\eta)}{4\eta^{1/4}\lambda},
\qquad \langle\tilde\phi_3\rangle=-\frac{m^2 \sqrt{\xi}}{4 \lambda},\qquad
 V_*=
\frac{m^4}{16\lambda^2 }
\eea

\medskip
\noindent The vev's of $\tilde\phi_{2,3}$ vanish  when $\xi\,m^2\ra
0$. In terms of the  component fields of the original Lagrangian
(\ref{originalL}),  the  values (\ref{rtq}) correspond to
$F={m^2}/{4 \,\lambda}$. Therefore, as expected $V_*=F^2$; this is a
check of above results since the value  of the potential at an
extremum point does not depend on the dynamical nature of the
original F. These relations are easily seen to be consistent with
the relation between $\tilde\Phi_D,\tilde\Phi$ and  new
$\tilde\Phi_i$ see (\ref{eee}) (and also (\ref{trasf}), (\ref{sr})).

Next we compute
the matrix of second derivatives of $V$ and
its eigenvalues  at this extremum point.
As discussed, we
use a diagonalisation method which takes into account  the
negative signature of the kinetic terms of the two ghosts fields.
We find
\medskip
\bea\label{newbasis}
m^2_{\tilde{\tilde\phi}_1}&=& \frac{-m^2}{1+\sqrt{1+2\xi\,m^2}}
\qquad \quad\qquad \quad
m^2_{\tilde{\tilde\phi}_2} =\frac{m^2}{1+\sqrt{1-2\,\xi\,m^2}}
\nonumber\\[10pt]
m^2_{\tilde{\tilde\phi}_3}&=&  \frac{m^2}{-1+\sqrt{1+2\,\xi\,m^2}}
,\qquad\quad\quad
m^2_{\tilde{\tilde\phi}_4}= \frac{m^2}{1-\sqrt{1-2\,\xi\,m^2}},\qquad\quad
m^2_{\tilde{\tilde\phi}_{5,6}}=\frac{1}{\xi}
\eea

\medskip
\noindent
The values $m^2_{\tilde{\tilde\phi}_{3,4,5,6}}$ are now  all positive,  and do
not suggest a local instability. The signs of
$m^2_{\tilde{\tilde\phi}_{1,2}}\approx \mp m^2/2$
are independent of the rescaling of the ghosts real component fields.
Finally,  $m_{\tilde{\tilde\phi}_{1,2}}^2$
are counterparts to those at the saddle point
of $V$  in the absence of the higher derivative term,
 given by $\pm m^2/2$, for same corresponding vev of
$\tilde\phi_{1,2}$ and $\phi$ respectively (see section~\ref{WZnohdo}).
Due to non-zero vev's of the auxiliary fields,
supersymmetry is in this case broken, similarly to Wess-Zumino model
without higher dimensional operators.
This ends our discussion on the spectrum  obtained from the second
order Lagrangian.

There remains the question of the relation between the potential $V$ in
(\ref{pot4})  and that of the original theory (\ref{originalL})
and how the latter is recovered from the former in the
limit\footnote{Eq.(\ref{rescale1}) is
singular if $\xi\ra 0$,  $\nu_{2,3}\ra 0$; in 
(\ref{pot4})  $m_{13}^2-m_{23}^2=-1/\xi$,
$m_{13}+m_{23}\sim - 1/\sqrt\xi$ are singular too.} $\xi\ra 0$.
To this purpose, evaluate  $V$ of  (\ref{pot4})  for  extremum
vev's given in eqs.(\ref{re1}), (\ref{re2}) but not in
(\ref{re3}). The  value obtained is
\medskip
\bea\label{rrrr}
V_{*}'&=&
\sqrt{\eta}\,\,\Big\vert \tilde\lambda_{111} \,\langle\tilde
\phi_-\rangle^2
+\tilde m_{11}\,\,\langle\tilde \phi_-\rangle
\,\Big\vert^2
=
\Big\vert \,\lambda \,\eta^{-\frac{1}{2}}\langle\tilde \phi_-\rangle^2
-m\,\,\eta^{-\frac{1}{4}}
\langle\tilde \phi_-\rangle\,\Big\vert^2
\eea

\medskip\noindent
 with $\tilde\phi_-\equiv
 \tilde\phi_1-\tilde\phi_2$. $V_*'$ is thus the value of $V$ evaluated at
extremum vev's in  directions other than $\tilde\phi_1-\tilde\phi_2$.
One observes that the extremum condition on $V_{*}'$
with respect to ``variable''
$\langle\tilde\phi_-\rangle\equiv \langle\tilde\phi_1-\tilde\phi_2\rangle$
recovers the  remaining condition  eq.(\ref{re3})
of  those in (\ref{re1}) to (\ref{re3}). On $V_{*}'$ the
limit $\xi\ra 0$ is well defined and finite.
Recalling the  scalar potential of the original theory
\bea
V=\Big\vert \lambda\,\phi^2 + m\,\phi\,\Big\vert^2,\qquad{\rm and}\qquad
\phi=-\frac{1}{\eta^{1/4}}\,(\tilde\phi_1-\tilde\phi_2)
\eea
one recovers eqs.(\ref{rrrr}).
To conclude, the scalar potential  in the original theory
(\ref{originalL}) is a ``projection'' of a more general potential
which includes the extra (ghost) degrees of freedom introduced by the higher
derivative term in the action, evaluated for extremum  vev's
of two linear combinations of {\it all} degrees of freedom.
 That is, there is no clear separation in
the potential between particle and ghost directions.
This is not too surprising if we recall  eq.(\ref{eee})
showing the  original $\Phi$ had itself a ghost ``piece''.

\subsection{The case of a general superpotential.}

We  extend the results of the previous section
 to the case of an
arbitrary superpotential $W(\Phi,S)$,
which can include higher dimensional operators,
but no higher derivative terms\footnote{This case is discussed in
  Section~\ref{sec3}.}.
One can also have additional superfields,
generically denoted here $S$, with standard kinetic
terms\footnote{If $S$ itself has higher derivative  kinetic terms
one introduces extra constraint superfields see previous section.}.
Consider
\medskip
\begin{eqnarray}\label{LL6}
\cL&=&\int d^4 \theta\, \Big[
\Phi^\dagger \Big(1+\,\xi\,\Box\Big) \Phi +
S^\dagger S\,\Big]
+\bigg\{ \int d^2 \theta \,\,\, W(\Phi,S)
 +h.c.\bigg\}
\eea

\medskip
\noindent
Following  steps similar to those in
 the previous section, $\cL$ is  shown to
be equivalent to
\medskip
\bea
\cL'&=&
\int d^4 \theta\,\,
\Big[
\tilde\Phi_1^{\dagger} \tilde\Phi_1
-\tilde\Phi_2^{\dagger} \tilde \Phi_2
-\tilde\Phi_3^{\dagger}\tilde\Phi_3
+S^\dagger S\,
\Big]\nonumber
\\[12pt]
&+&
\bigg\{
\int d^2 \theta \,\bigg[
\frac{1}{2}\,
\tilde \mu_{kp}\,\tilde\Phi_k\,\tilde\Phi_p +
W[\Phi(\tilde\Phi_{1,2,3}),\,S]\bigg]+h.c.
\bigg\},\label{genW}
\eea

\medskip
\noindent
 with the following relations:
 \medskip
 \bea
 \Phi(\tilde\Phi_{1,2,3})=\frac{1}{\eta^{1/4}}
 \Big[(\tilde\Phi_2-\tilde\Phi_1)\,\Big],\qquad \quad
 \Phi_D=\frac{4}{m\,\sqrt\xi}\,\tilde\Phi_3
 \label{phiphid}
\eea

\medskip
\noindent
Notice that $S$ is spectator under going from the fourth order to
second order theory, since it has no higher derivative kinetic terms
and does not mix with the kinetic terms of $\Phi$.
One also finds that
 $\tilde\mu_{ij}=0$
for all $i,j$ except:
\medskip
\bea\label{mu2}
\tilde \mu_{13}=\tilde\mu_{31}= \frac{1-\sqrt{\eta}}{2\,\eta^{1/4}\,\sqrt\xi},
\qquad\,\qquad
\tilde \mu_{23}=\tilde\mu_{32}=
-\frac{1+\sqrt\eta}{2\,\eta^{1/4}\,\sqrt\xi}
\eea

\medskip
\noindent
where as usual $\eta=1+4\xi\,m^2$. These values are
equal to  $\tilde m_{13}, \tilde m_{23}$ of
(\ref{newmass}) and are generated by the constraint (\ref{delta-L})
alone, and not by
 bilinears that may be present in $W$. 
Also $\tilde\mu_{13}=\tilde\mu_{31}\approx 0$,
$\tilde\mu_{23}=\tilde\mu_{32}\approx -1/\sqrt\xi$ for 
 $\xi\,m^2\ll 1$.
 The auxiliary fields are
\medskip
\bea
\tilde
F_1^*&=&-\tilde\mu_{31}\,\tilde\phi_3 +
{\eta^{-1/4}}\,\, W'_\phi\nonumber\\[10pt]
-F_2^*&=&-\tilde\mu_{32}\,\tilde\phi_3 -
{\eta^{-1/4}}\,\, W_\phi'  \nonumber\\[10pt]
-F_3^*&=&-\tilde\mu_{k3}\,\tilde\phi_k,\qquad F_S^*=
-  W'_{\phi_s}
\label{fs}
\eea

\medskip
\noindent
 Here $W_\phi'\equiv \partial W(\phi,\phi_S)/\partial\phi$, with
 $\phi$ replaced by $\phi=\phi(\tilde\phi_{1,2})$
of (\ref{phiphid}).
The scalar potential becomes
\smallskip
\bea\label{ff}
V= \Big\vert\, \tilde\mu_{31} \,
\tilde\phi_3- {\eta^{-1/4}}
\,W_\phi'
\,\Big\vert^2
-
\Big\vert\, \tilde\mu_{32}\, \tilde\phi_3+ {\eta^{-1/4}} \,W_\phi' \,\Big\vert^2
-\Big\vert \tilde\mu_{k3}\,\tilde\phi_k\Big\vert^2
+\Big\vert  W_{\phi_s}' \,\Big\vert^2
\eea

\medskip
\noindent
In the original $\Phi$-dependent formulation, the potential
which was function of $F, \phi,F_s,\phi_s$, evaluated at an extremum
point(s) labelled by ``o'', was
\bea
V_o=\Big[\big\vert\, W'_\phi\,\big\vert^2 +\big\vert\,
  W'_{\phi_s}\,\big\vert^2
\Big]_{o}\eea
Let us  assume that supersymmetry is unbroken.
In the original-field language this means that
 $W'_\phi=W'_{\phi_s}=0$ and also $F=F_s=0$ {\it at this extremum
  point}.
 This is true regardless of the dynamical nature of $F$.
Let us now  investigate the corresponding
 situation in the second order theory.
The extremum conditions for (\ref{ff}),
${\partial V}/{\partial \tilde\phi_p}=0$, $p=1,2,3$ give,
at the extremum point considered:
%\medskip
\bea\label{eqsol}
\langle\tilde\phi_3\rangle=0,\qquad \langle\tilde\mu_{3k}
\,\tilde\phi_k\rangle=0,\quad k=1,2.
\eea
From (\ref{fs}),  at the extremum point of the scalar potential
we find $\tilde F_i=F_s=0$, $i=1,2,3$.
 The vanishing of the auxiliary fields of the second order theory
 confirms, in the new formalism, that supersymmetry is
unbroken in this state, as the equivalence of the two formulations
of the theory would suggest.

\section{Effects of higher derivatives in the superpotential}
\label{sec3}

The method developed so far can  be applied
when higher derivative terms are  present in the superpotential.
Consider  the  Lagrangian

\begin{eqnarray}\label{oW}
\cL&=&\int d^4 \theta\,\, \Phi^\dagger  \Phi +
\bigg\{
\int d^2 \theta \,\bigg[s_2 \,\sqrt{\xi}\,\, \Phi\,\,\Box \,\,\Phi +
\frac{m}{2}\,\, \Phi^2 +
\frac{\lambda}{3}\,\, \Phi^3\bigg] +h.c.\bigg\}
 \nonumber\\[12pt]
&=&
\int d^4 \theta\,\, \Big[
\Phi^\dagger \Phi
+ \frac{s_2\,\sqrt{\xi}}{4} \Big(
\Phi\, D^2 \,\Phi +\Phi^\dagger \,{\overline D}^2\,\Phi^\dagger\Big)
\Big]
+ \bigg\{\int d^2 \theta \,\Big[\frac{m}{2}\, \Phi^2 +
\frac{\lambda}{3}\, \Phi^3\Big] +h.c.\bigg\}\quad
\end{eqnarray}

\medskip
\noindent
where\footnote{We used $\int d^4x\, d^2\theta \,\,\overline D^2
 Q(x,\theta,\bar\theta)=-4\int  d^4 x\,  d^4\theta
\,\,Q(x,\theta,\bar\theta)$.} we allow $s_2=\pm 1$.
We follow  the steps of the previous section, introduce
 (\ref{trasf}),  a  Lagrange multiplier chiral
superfield and $\Delta \cL$, eqs.(\ref{constraint}),
 (\ref{delta-L}).
The counterpart to eqs.(\ref{roots1}) has now the roots
\medskip
\bea\label{swq}
\nu_{1,2}
=\frac{1}{2}\big[\,1\pm \sqrt\eta\prime\,\big],
\qquad
\nu_3=0,\qquad{\rm where}\qquad \eta^\prime\equiv 1+ 4 m^2\,\xi\,(1+s_2^2/16)
\eea

\medskip
\noindent
with the choice $\nu_1>0$. Unlike in eq.(\ref{roots2}), there is a
vanishing eigenvalue, so  there is one ghost
superfield and one particle-like superfield.  After appropriate
normalisation of the Kahler terms, the Lagrangian equivalent to
that in (\ref{oW}) is:
\medskip
\begin{eqnarray}
\cL'&=& \int d^4 \theta\,\,
\bigg[\tilde\Phi_1^{\dagger} \tilde\Phi_1
-\tilde\Phi_2^{\dagger} \tilde\Phi_2 \bigg]
+\bigg\{
\int d^2 \theta \,\bigg[\frac{1}{2}\, \tilde m_{kp} \,\tilde\Phi_k \,\tilde\Phi_p
+\frac{1}{3}\,\tilde\lambda_{kpl}\,\tilde\Phi_k\,\tilde\Phi_p\,\tilde\Phi_l
\bigg] +h.c.\bigg\},\qquad
\eea

\medskip
\noindent
where
\bea
\tilde m_{kp}=\frac{m_{kp}}{\sqrt{\vert \nu_k^{q_{k3}} \,\nu_p^{q_{p3}}\,\vert}},
\qquad
\tilde\lambda_{kpl}=\frac{\lambda_{kpl}}{\sqrt{\vert \nu_k^{q_{k3}}
    \,\nu_p^{q_{p3}}\,\nu_l^{q_{l3}}\vert}},\qquad k,p,l=1,2,3.
\eea

\medskip
\noindent
with $q_{k3}=1-\delta_{k3}$,
and $\tilde m_{kp}$, $\tilde\lambda_{kpl}$
are given in  (\ref{newmasses}) with $v_{ij}$ presented in the
Appendix, eq.(\ref{spthdo}). As before $\tilde\lambda_{ij3}=0$
$i,j=1,2,3$.
$\tilde\Phi_3$ can be eliminated using the equations of motion:
\medskip
\bea
\tilde m_{k3}\,\tilde \Phi_k=0,\qquad\quad
\tilde\Phi_3=-\frac{1}{\tilde m_{33}}\,\big(\tilde m_{13}
\,\tilde\Phi_1+\tilde m_{23}\,\tilde\Phi_2\big)
\eea
Unlike in Section~\ref{sec1}, here $\tilde\Phi_3$ can be eliminated, 
and this is ultimately due to the fact that in (\ref{oW}) $F$ is not
dynamical and can be integrated out.
As a result  $\cL'$ can be re-written 
\medskip
\begin{eqnarray}
\cL'&=&
\int d^4 \theta\,\,
\bigg[
\tilde\Phi_1^{\dagger} \tilde\Phi_1
-
\tilde\Phi_2^{\dagger} \tilde\Phi_2
\bigg] \nonumber\\[12pt]
&\!\!\!\!\!\!\!\!\!\!\!\!\!\!+&
\!\!\!\!\!\!\!\!\!\!\!
\bigg\{
\int d^2 \theta \,\bigg[
 \frac{1}{2}\,\big(\,d_1\, \tilde\Phi_1^2 +d_2\,\tilde\Phi_2^2 +2 \,d_3\,
 \tilde\Phi_1\tilde\Phi_2\big)
+
\frac{\tilde\mu}{2}\,\big(\tilde\Phi_1-\tilde\Phi_2\big)^2+
\frac{\tilde\lambda_{111}}{3}\,\big(\tilde\Phi_1-\tilde\Phi_2\big)^3
\bigg]+h.c.\bigg\}\qquad
\eea

\medskip
\noindent
where
\bea\label{mm1}
  \tilde\mu=\frac{m}{\sqrt{\eta^\prime}},
  \qquad
  \tilde\lambda_{111}=-\frac{\lambda}{\eta^{\prime\,3/4}},
\eea
and
\bea\label{mm2}
d_1=
\frac{(\sqrt{\eta^\prime}-1)^2}{8\,s_2\sqrt{\eta^\prime\,\xi}},
\qquad
d_2=
\frac{(\sqrt{\eta^\prime}+1)^2}{8\,s_2\sqrt{\eta^\prime\,\xi}},
\qquad
d_3=
\frac{{\eta^\prime}-1}{8\,s_2\sqrt{\eta^\prime\,\xi}}
\eea

\medskip
\noindent
with $d_1=\cO(\xi^{3/2} m^4)$ and  $d_3=\cO(\sqrt\xi\,m^2)$ vanishing
in the limit of small $\xi\,m^2$;  finally
$d_2=\cO(1/\sqrt\xi)$ gives the leading contribution to the mass of
the  ghost superfield \footnote{The whole field-dependent term
 involving coefficients $d_{1,2,3}$ equals to $(s_2/16)\,\sqrt\xi \,\Phi_D^2$
with $\Phi_D$ of (\ref{oldandnew}), (\ref{trasf}).}
$\tilde\Phi_2$.

One also finds the relation between old and new fields:
\medskip
\bea\label{oldandnew}
\Phi=\frac{1}{\eta^{\prime\,1/4}}\,
\big[\tilde\Phi_2-\tilde\Phi_1\big],\qquad
\quad
\Phi_D=\frac{1}{\eta^{\prime 1/4}\,\sqrt\xi\,s_2}
\,\Big[ \, (1-\sqrt{\eta^\prime}\,)\,\tilde\Phi_1
-(1+\sqrt{\eta^\prime}\,)\,\tilde\Phi_2\Big]
\eea

\medskip
\noindent
with $\Phi_D$ introduced  in (\ref{trasf}). Eqs.(\ref{mm1}), (\ref{mm2})
show again that the couplings of the second order theory acquired a
scale ($\xi$) dependence, via fields rescaling, first eq in  (\ref{oldandnew}).

The scalar potential is
\smallskip
\bea\label{tv}
V&=& \vert \tilde F_1\vert^2 - \vert \tilde F_2\vert^2
\nonumber\\[10pt]
&=&
\big\vert\,
\tilde\mu\,
(\tilde\phi_1-\tilde\phi_2)+\tilde\lambda_{111}\,
(\tilde\phi_1-\tilde\phi_2)^2
+d_1\,\tilde\phi_1+d_3\,\tilde\phi_2\,\big\vert^2
\nonumber\\[10pt]
&-&
\big\vert\,
\tilde\mu\,(\tilde\phi_1-\tilde\phi_2)+\tilde\lambda_{111}\,
(\tilde\phi_1-\tilde\phi_2)^2
-d_3\,\tilde\phi_1-d_2\,\tilde\phi_2\,\big\vert^2
\eea

\medskip
\noindent
In the basis of the second order theory, the potential
is not positive definite anymore, similarly to the previous
section.
There are two ground states,
for
\medskip
 \bea
i)\quad \langle\tilde\phi_1\rangle&=&\langle\tilde\phi_2
\rangle=0\nonumber\\[12pt]
ii)\quad
\langle\tilde\phi_1\rangle&=&
\frac{m\,(1+\sqrt{\eta^\prime})}
{2\,\eta^{\prime\,1/4}\,\lambda},
\qquad\qquad
\langle\tilde\phi_2\rangle=
\frac{m\,(1-\sqrt{\eta^\prime})}
{2\,\eta^{\prime\,1/4} \,\lambda}
\eea

\medskip
\noindent
which are similar to their counterparts in (\ref{casei}), (\ref{caseii}).
The vev's above give that
$(\langle\tilde\phi_2\rangle-\langle\tilde\phi_1\rangle)/\eta^{\prime
  1/4}$ equals  $0$ for $i)$,  and $ -m/\lambda$ for $ii)$.
This result for the two ground states
 is in agreement with what one obtains in the Wess-Zumino model
in the absence of higher derivative operators, using the first
relation in (\ref{oldandnew}), for the corresponding
ground states $\langle\phi\rangle=0$ and $-m/\lambda$, see
section~\ref{WZnohdo}. In the limit of decoupling the higher
dimensional operator $\xi\ra 0$ then $\eta^\prime\ra 1$,
$\langle\tilde\phi_2\rangle=0$ and
$-\langle\tilde\phi_1\rangle\ra \langle\phi\rangle$.
 The ghost (super)field decouples and one
recovers the Wess-Zumino model without higher derivatives.
For both $i)$ and $ii)$ cases:
\bea
\langle \tilde F_1\rangle=\langle \tilde F_2\rangle=0,\qquad V_*=0
\eea

\medskip
\noindent
i.e. supersymmetry is unbroken, in agreement with the picture
in the original basis for the corresponding ground states.
One  then computes the spectrum for
i), ii), in basis  $\tilde\phi_{1,2}$
with a  metric which takes account of the negative sign of
the kinetic term of $\tilde\Phi_2$, similar to the previous section. The
solutions are (with fixed $s_2=\pm 1$):

\bea\label{wqwq}
m_{\tilde{\tilde{\phi}}_{1,2}}^2=
\frac{1}{8\,\xi}\Big[1\mp
  \sqrt{1+8\,m\,s_2\sqrt\xi}+4\,m\,s_2\,\sqrt\xi\Big] \ .
\eea

\medskip
\noindent
These values  agree with those obtained from the  poles of scalar
propagators found using the old
basis
 ($\phi, F$) after
performing the Grassmann integrals in  first line of
(\ref{oW}). The above values
are of order  $m^2+\cO(m^3\,\sqrt\xi)$ for $\tilde{\tilde\phi}_1$ and
$1/(4\xi)+\cO(m/\sqrt\xi)$ for the
ghost $(\tilde{\tilde\phi}_2)$.
Note that the  correction to the mass of
$\tilde{\tilde\phi}_1$ is suppressed only by $1/M_*$ and is thus 
larger  than the one discussed in the case of Kahler higher derivative
terms, eq.(\ref{w1p}) suppressed by $\xi=1/M_*^2$.
The  effects of the operator $\Phi\Box \Phi$ for  phenomenology are
discussed in Section~\ref{sec6}.

\subsection{The case of a general superpotential.}

The previous analysis is easily extended  to
an arbitrary superpotential  in addition to
the higher dimensional (derivative) term. Consider the Lagrangian
\medskip
\bea\label{gW}
\cL&=&\int d^4 \theta\, \Big[\,\Phi^\dagger  \Phi +S^\dagger\,S\,\Big]+
\bigg\{
\int d^2 \theta \,\Big[s_2 \,\sqrt{\xi}\,\, \Phi\,\,\Box\, \Phi +
W(\Phi,S)\Big] +h.c.\bigg\}
\eea

\medskip
\noindent
Here $W(\Phi,S)$ has no  derivative terms in any of the fields,
 but otherwise is arbitrary\footnote{If higher derivative
terms in $S$ exist in the superpotential, the same method is also applied
 for $S$.}, and can include non-renormalisable
 interactions. $S$ is an arbitrary superfield.
One shows that the  Lagrangian equivalent to $\cL$ is:
\medskip
\begin{eqnarray}
\cL'&=& \int d^4 \theta\,\,
\Big[\tilde\Phi_1^{\dagger} \tilde\Phi_1-
\tilde\Phi_2^{\dagger} \tilde\Phi_2 +S^\dagger\,S\, \Big] \nonumber\\[12pt]
&+& \bigg\{ \int d^2 \theta \,\,\bigg[\,
\, \frac{1}{2}\,\big(\,d_1\, \tilde\Phi_1^2 +d_2\,\tilde\Phi_2^2 +2 \,d_3\,
 \tilde\Phi_1\tilde\Phi_2\big)
+W(\Phi(\tilde\Phi_{1,2}),S)\, \bigg]+h.c.\bigg\}\qquad
\eea

\medskip
\noindent
The coefficients $d_{1,2,3}$ are given in
 eqs.(\ref{mm2}) and the relation between old and new fields
is that of (\ref{oldandnew}) which applies in
this case too. Finally, $S$ is spectator under
the unfolding of the fourth order theory into the
second order one.
The auxiliary fields are
\medskip
\bea
-\tilde F_1^* &=& d_1\,\tilde\phi_1+d_3\,\tilde\phi_2-\eta^{\prime
  -1/4}\,W^\prime_\phi
\nonumber\\[10pt]
\tilde F_2^* &=& d_2\,\tilde\phi_2+d_3\,\tilde\phi_1+\eta^{\prime -1/4}
\,W_\phi^\prime,\qquad
- F_s^*= W^\prime_{\phi_s}\,
\eea

\medskip
\noindent
where $W'_\phi$ ($W'_{\phi_s}$) is the partial derivative wrt $\phi$\,
($\phi_s$).
Then the scalar potential is equal to
$
V=\vert\, \tilde F_1\,\vert^2
-\vert\, \tilde F_2\,\vert^2
+\vert\, \tilde F_s\,\vert^2
$.
Assume now that our model is in a ground state
having (in the old basis) $F=F_s=0$ at the extremum point
of the potential i.e.
supersymmetry is unbroken. To picture this  in the new formalism, use
eq.(\ref{oldandnew}) giving
 $\tilde F_1=\tilde F_2$. Further, the extremum conditions of the scalar
potential wrt the new basis give three eqs which depend on
$W_{\phi}'$ and $W_{\phi_s}'$ and on the second derivatives of the
superpotential wrt $\phi$, $\phi_s$, evaluated at the
extremum  point considered.  One also uses that
 $W_{\phi}'=0$ and $W_{\phi_s}'=0$
at the extremum point, while the second derivatives can be non-zero
for this state. With these observations, one immediately finds that $\tilde
 F_1=\tilde F_2=0$ which  recovers, in the new field basis
that supersymmetry is  unbroken, as expected by the equivalence
of the two formulations.

The analysis can in principle be extended to the case when
 higher derivative terms of type discussed in sections~\ref{hdosec},
\ref{sec3} are {\it simultaneously} present
in the Kahler term and the superpotential,
for an otherwise  arbitrary superpotential. The method can
also  be applied to  terms such as  $\Phi^n\Box\,\Phi$
in the superpotential or $(\Phi^\dagger)^2\,\Phi+h.c.$ etc,
in the Kahler part of the action.

\section{Supersymmetry breaking and higher-derivatives }\label{sec4}

\subsection{A model of supersymmetry breaking}\label{SSB}

A natural question, which was our main motivation in studying
theories with higher dimensional operators, is whether
supersymmetry can be spontaneously broken due to the higher derivative 
terms, or equivalently in the two-derivative formulation, if 
supersymmetry breaking can be triggered by the presence of the 
ghost field(s).

The purpose of this section is to show the 
importance of   the relation between the two formulations 
of a  theory with higher derivatives found in the previous sections,
for the case of  supersymmetry breaking.
For example one can have  models with higher derivative terms
which look rather uninteresting in the original (higher derivative)
formulation and could  be disregarded when decoupling the higher
derivative term, and which in the two-derivative formulation 
are actually interacting theories and exhibit (spontaneous)  
supersymmetry breaking.

Here we provide an example of  a model  which
in the limit of vanishing higher derivative operator has a trivial
SUSY breaking, in the sense that the theory becomes free with a
positive cosmological constant. However, in the presence of the higher
derivative operator and in the second order formulation,
the theory is interacting and has spontaneous supersymmetry breaking 
\`a la O'Raifeartaigh \cite{o'r}. 
The example we consider starts from the two-derivative
formulation with one particle (super)fields $S$ and two ghost
superfields $\Phi, \chi$
\medskip
 \bea \cL \ = \ \int d^4 \theta 
\left(S^\dagger \,S-\Phi^\dagger\,\Phi-\chi^{\dagger} \chi \right) \ 
+ \ \int d^2 \theta \left[ S\,(- \lambda \chi^2 - m^2) - M_* \Phi \chi  +
  {\rm h.c.}  \right] \ . \label{sb1}
\eea 

\medskip
\noindent
The theory breaks SUSY \`a la O'Raifeartaigh since the SUSY conditions 
\bea
- F^*_S \ = \ -\lambda \,\phi_{\chi}^2 - m^2 = 0 \quad , 
\quad 
- F^*_\phi\ = \ M_* \,\phi_{\chi} = 0  \ 
\label{sb2} 
\eea
cannot be simultaneously satisfied.
The vacuum of (\ref{sb1}) is given by 
\bea \langle \phi_{\chi} \rangle =
\langle \phi_{\Phi} \rangle = 0 \quad , \quad \langle 
\phi_S \rangle =
v_2 = {\rm arbitrary} \ , \label{sb01} 
\eea 

\medskip
\noindent
therefore the scale of
SUSY breaking is given by $F^*_S \ = \ m^2$.
 In the limit of large ghost mass $M_* \gg m$, we can replace
$\chi$ by the classical superfield eq. 
\bea \chi \ = \ \frac{1}{4 M_*} {\bar D}^2 \Phi^{\dagger} \ . 
\label{sb3} 
\eea 
Inserting (\ref{sb3}) back
 into the original action (\ref{sb1}) we find, after some
standard manipulations, the Lagrangian\footnote{One can 
start in (\ref{sb1}) with $+\Phi^\dagger \Phi$, then in
(\ref{sb4}) $\Phi^\dagger \Box\Phi$ has opposite sign, see also 
(\ref{originalL}), (\ref{roots2}), (\ref{LL2}).}
\medskip
\bea \cL & = & \ \int d^4
\theta \left[ S^{\dagger} S + \Phi^{\dagger} \Phi + 
\frac{1}{M_*^2}  \Phi^{\dagger} \Box \Phi + \frac{\lambda}{4 M_*^2}
\,(S \Phi^{\dagger} {\bar D}^2 \Phi^{\dagger} 
+h.c.) 
\right] \nonumber\\[10pt]
 &+& 
\bigg[\int d^2 \theta \left( - m^2 S\right) + {\rm h.c.}\bigg] \ .
\label{sb4} \eea 

\medskip
\noindent
Notice that it is safe to replace $\chi$ by
(\ref{sb3}) since in the original theory $\chi$ did not contribute
to SUSY breaking. Notice also the sign flip in the kinetic term of
$\Phi$, which became a standard kinetic term, supplemented by the
two higher  derivative operators. Of these operators
 the first one was  considered already
in Section~\ref{sec1}, whereas the second 
one is of the form (d) of eq.(\ref{derivs}),  not considered before.  

In the decoupling limit $M_* \rightarrow \infty$ eq.(\ref{sb4}) describes a free
supersymmetric theory for the two fields $S,\,\Phi$ with a linear
superpotential $W = - m^2 \,S$ which breaks supersymmetry in a
trivial way, by adding a pure cosmological constant. Switching on
 the higher derivative terms generates an interacting theory whose
 SUSY breaking
can be better described in the two-derivative version as an O'Raifeartaigh model
(\ref{sb1}). For $M_* ^2 \gg m^2$, both $\chi$ and $\Phi$
are in fact very heavy, whereas $S$ remains massless. We could
have therefore integrated out $\Phi$ instead, in which case however
we would turn $\chi$ into a particle-like superfield. It would be more
satisfactory to integrate both $\chi,\Phi$ simultaneously. In this
case, however, the  theory for $S$ is non-local and highly non-linear.

Can we use the method
developed in the previous sections to go from eq.(\ref{sb4})
to (\ref{sb1})? The answer is indeed affirmative, despite the presence
of the new term proportional to $\lambda$ in the Kahler term (not
present before). This can be
easily checked using eqs.(\ref{trasf}), (\ref{genW}), (\ref{phiphid}),
(\ref{mu2}), for $\xi= 1/M_*^2$.
The term in (\ref{sb4}) proportional to $\lambda$ can be ``moved'' into the
superpotential where it acquires an extra  $\overline D^2$ 
and becomes  of type $S\,(\overline D^2\Phi^\dagger)^2$ (using that
$S$ is chiral), and which upon using second eq in (\ref{trasf}) and
 (\ref{phiphid})
 becomes a {\it non-derivative} interaction  term. This interaction term
 recovers the first term in the superpotential in (\ref{sb1}). One
 then uses that $\tilde\mu_{13}$ vanishes in the limit $\xi\ra 0$ while
$\tilde\mu_{23}\ra -{1}/{\sqrt\xi}$. The latter will in the end
recover the last term in the superpotential of (\ref{sb1}) (see also (\ref{genW})).
Finally, the first three  D-terms in (\ref{sb4}) become, 
using (\ref{trasf}), the D-terms of (\ref{sb1}) after disregarding 
the Kahler term of a non-interacting, massless superfield.
This concludes our discussion on how to recover from (\ref{sb4}),
eq.(\ref{sb1}).

It is important  to notice that the formalism of previous sections
applies not only
in the presence of Gaussian-like terms (as it would be inferred
 from the discussion in Sections~\ref{sec1},\ref{sec3})
 but also for other  terms, like the last D-term in
(\ref{sb4}). Finally, the method can be iterated for  models 
with an even larger number of derivatives, to map it
to a two-derivative theory. As a result the latter  may then acquire 
higher dimensional superpotential interactions\footnote{for a
  sufficiently large number of derivatives} but no
 higher derivatives.

\subsection{Soft breaking terms}

We return  to the  models of Sections~\ref{sec1},\ref{sec3}
of  eqs.(\ref{originalL}), (\ref{oW})
(or more generally eqs.(\ref{LL6}), (\ref{gW})\footnote{In this case 
soft terms in $S$ in addition to those below can also be present.}),
to comment on supersymmetry  breaking. The results below apply to
both of these models as we shall see shortly.
Assuming  that supersymmetry is broken by  explicit soft terms 
\cite{GG} added to (\ref{originalL}) and (\ref{oW}) respectively,   
let us investigate their explicit form in their second order,
equivalent  formulation. 
Consider therefore the addition of $\cL_{soft}(\phi,\phi^*)$ to 
eq.(\ref{originalL}), (\ref{oW}) where 
\medskip
\bea -
\cL_{soft}=M_1^2\,\phi^*\,\phi+ (M^2_2 \,\phi^2+h.c) 
\eea 

\medskip
\noindent
Taking
into account  the relation between $\phi$ and $\tilde\phi_i$ which
is similar for  Sections \ref{sec1} and \ref{sec3},
 see (\ref{phiphid}), (\ref{oldandnew}),
 the soft terms become
\medskip
\bea -\cL_{soft}= \frac{M_1^2}{\sqrt{\beta}}\,\,\Big\vert
\tilde\phi_1-\tilde\phi_2\Big\vert^2+
\bigg[\frac{M_2^2}{\sqrt{\beta}}\,\,(\tilde\phi_1-\tilde\phi_2)^2+c.c.\bigg],
\eea where $\beta$ is equal to: \bea
 \eta&\equiv&
1+4\,\xi\,m^2 \qquad\qquad\quad \qquad{\rm (for\,\,\,
 Section~\ref{hdosec})}\nonumber\\[10pt]
 \eta'&\equiv&1+(17/4)\,\xi\,m^2\,
\qquad\qquad\quad {\rm (for\,\,\,Section~\ref{sec3})} \eea

\medskip
\noindent which are thus of identical form up to a rescaling of
$\xi$. Similar relations apply for trilinear  soft terms. The soft
breaking terms also acquired a scale/moduli dependence on
$1/\xi=M_*^2$ which is the scale of the higher derivative operators.
This dependence is introduced dynamically by the ``constraint''
Lagrangian of eq.(\ref{delta-L}).

It is important to mention here that the presence of soft terms
does not affect the holomorphic constraint and that
the formalism we developed in previous sections is not affected. 
We checked this for specific cases
by computing  the spectrum after adding the
soft terms, in both formulations (with 4- and 2-derivatives).
In the second order formulation this was done using the eigenvalues 
of the second
derivatives matrix of the potential with an appropriate metric
in the field space, as detailed in previous sections.

\subsection{Further remarks on supersymmetry breaking}

We end this discussion with  more general remarks on models with ghost
superfields in  the second order  action.  In these, the
 scalar potential is of the generic form
\bea V=\sum_{i} \vert F_i\vert^2-\sum_{j}\vert F_j\vert^2. \eea
 were the first sum accounts for contributions from
particles and the second for that of the ghosts superfields present
in the model considered. One could  in principle have $V>0$, $V<0$
or even $V=0$ with  broken supersymmetry. The breaking can in
principle be done
by vev's of the particle-like $F_i\not= 0$, by the ghost-like
$F_j\not=0$'s  or by both.
 For example a toy model with
\medskip
\bea \cL_1=\int d^4\theta \,\Big[\Phi_1^\dagger\,\Phi_1 -
\Phi_2^\dagger\,\Phi_2\Big] +\bigg[\int d^2\theta \,
W(\Phi_1-\Phi_2)+ h.c.\bigg] +m_0^2\,(\phi_1-\phi_2)^2 \eea can have
a vanishing scalar potential, with broken supersymmetry. Indeed, the
two auxiliary fields have identical eqs of motion, and thus
$V(\phi_1,\phi_2)=V_{soft}(\phi_1,\phi_2)$ so $V$ has a minimum at
$\phi_1-\phi_2=0$. Assuming  $W'(\phi_1-\phi_2)\not=0$ which is
easily satisfied if $W$ contains a linear term such as
$g(\phi_1-\phi_2)$, then supersymmetry is broken, $F_1=F_2=g\not=0$
yet the value of the scalar potential at the ground state is still
vanishing. This remark has some similarities to \cite{Kaplan:2005rr}
and may be  of  some interest for the cosmological constant problem
\cite{Weinberg}.

At this point one could raise the issue of the stability
\cite{ArkaniHamed:2003uy,Hawking} of the models with higher derivative
terms after supersymmetry breaking.  Some stability issues were discussed
in \cite{Hawking}, where it was shown that the transition probabilities
in such models have no exponential growth and  in the decoupling limit
($M_* \ra \infty$) tend to those in ordinary second order theories. The
price paid for stability is breaking unitarity at the scale $M_*$ which is
assumed to be very high relative to all other scales in the theory.

In the supersymmetric context of our models, further analysis 
of the stability is
necessary, along the lines discussed more recently in
\cite{ArkaniHamed:2003uy}, where the possibility of the formation 
of a ghost condensate was analysed in similar models.  
Additional constraints  \cite{ArkaniHamed:2003uy} were also derived from
imposing that the S-matrix respected all the standard axioms.
Our purpose was to illustrate the method of ``unfolding'' the theory with
higher dimensional operators (obtained for example after integrating
out massive states) into a second order theory; this method is general
and can be applied to models which eventually meet all the constraints
discussed in \cite{ArkaniHamed:2003uy}.
We believe that our second-order formalism is very useful for a
detailed analysis of stability, since it gives an off-shell description of the
dynamics of the system, whereas in the original four-order theory
supersymmetry is realized on-shell, since  auxiliary fields
did acquire their own dynamics.

\subsection{Renormalisability issues.}\label{ren}

Using our formalism we showed that a theory with higher derivative
operators of type considered in the previous sections, is  equivalent
to a theory without such operators but with additional superfields and
renormalised couplings. Such equivalence remained true in the
presence of soft breaking terms. If the initial theory 
had no other additional 
higher dimensional (non-derivative) operators, the equivalent second
order formulation has only dimension 4 operators. 
Such theory  can therefore be renormalisable. 
This is possible provided that we 
specify the analytical continuation of such theory to Euclidean 
metric. This is relevant since in models with ghosts the sign of the
$i\epsilon$ prescription in their propagators is very important for 
the UV behaviour of a softly broken theory and in some cases can dramatically 
alter it, see discussion in \cite{Antoniadis:2006pc} (despite a soft breaking and
contrary to what one would expect on naive dimensional
grounds\footnote{In the presence of higher derivative terms power
counting  for UV divergence of loop integrals does not
always work in  Minkowski space, for details see
 \cite{Antoniadis:2006pc}.}). However, if the 
propagators prescription for the ghost and particle-like degrees of freedom are
similar (i.e they both undergo Wick rotations in same sense),
then the Minkowski and Euclidean descriptions of the theory
have  similar  UV behaviour. In this case, the 2-derivative formulation of
the theory, which has only D=4 operators and is softly broken,
could actually be renormalisable\footnote{It would be useful to derive 
  such prescriptions from the original theory with higher derivative
  operators using  a path integral formulation in the Minkowski
  space-time. No such formulation is available  at present.}.

\section{Applications~: MSSM with higher-derivative
operators}\label{sec6}

As an application to the possible low-energy effect of
higher-derivative operators, we consider the Minimal Supersymmetric
Standard Model (MSSM) and the corrections to the Higgs mass coming
from such operators. We denote by $H_1,H_2$ the two MSSM Higgs doublets. The
relevant Lagrangian to consider is that of the MSSM supplemented by 
derivative 
operators built out of the Higgs superfields. The lowest dimensional ones
are\footnote{The effects of gauge interactions are not included in this section.}~: 
\bea \cL = \cL_{\rm MSSM} + \int d^4 \theta \left[ \xi_1
H_1^{\dagger} \Box H_1 + \xi_2 H_2^{\dagger} \Box H_2 \right] -
\left[ \int d^2 \theta \,\sqrt\xi_3 H_1 \Box H_2 + {\rm h.c.} \right] \ ,
\label{mssm1} \eea 
where $\cL_{\rm MSSM}$ is the standard MSSM
Lagrangian, including the soft-breaking terms. Notice first of all
that the higher derivative terms
do not change the vacuum structure of the theory. They
 change however the tree-level Higgs physical spectrum. Indeed,
by expanding the Lagrangian (\ref{mssm1}) around the vacuum breaking
the electroweak symmetry and restricting for simplicity to the case
$\xi_3=0$, we find the Lagrangian relevant for the scalar sector
\footnote{Here $h_i$ are the scalar Higgs components of the
superfields.} 
\medskip
\bea \cL^{(2)} = - h_{\bar i}^{\dagger} \left[
  \xi_i\,\Box^2\,+\Box\,\right] h_i
  +\xi_i\,F_i^*\,\Box\,F_i- V,\qquad\quad i=1,2
\label{mssm2} \eea 

\medskip
\noindent
where $V$ is that of the MSSM before eliminating the auxiliary 
$F_{1,2}$ of $H_{1,2}$.
One computes the corrected values of $m_h^2$, $m_H^2$ of the neutral scalar
eigenstates, using the poles of the corresponding propagators (vanishing
of the appropriate determinant in the basis  of Higgs and auxiliary
fields). Assuming  for  simplicity  that $\xi_1=\xi_2=\xi$, then one finds
\medskip
\bea 
2\, \xi\, p^4 -p^2 \,[1+(m_h^2-\mu^2)\,\xi]+ m_h^2 = 0 \quad , \quad  
2 \,\xi \,p^4 - p^2\,[1+(m_H^2-\mu^2)\,\xi]+ m_H^2 = 0 \ , \label{mssm3} \eea
 where $m_h$ and $m_H$ are the
masses computed in MSSM. Setting $p^2$ equal to the corrected
corresponding Higgs mass,
we find the leading corrections to the neutral Higgs masses to be 
\medskip
\bea 
\frac{\delta m_h^{2}}{  m_h^2} \ \approx
\ \xi \,( m_h^2+\mu^2)  \quad , \quad 
\frac{\delta m_H^{2}}{ m_H^2} \ \approx \ 
\xi\, (m_H^2+\mu^2) \ . \label{mssm4} 
\eea

\medskip
\noindent
 For a cutoff $M_*$ ($\xi_{1,2} =1/M_*^2$)
in the 5-10 TeV range, these effects are of order $1-2\%$
and therefore too small to give a sizable contribution.

 Let us now examine the effects of the operator
$W_1= \sqrt\xi_3 H_1 \Box H_2$ 
in the superpotential, where 
$\sqrt\xi_3=1/M_*^2$ and set $\xi_1=\xi_2=0$. It turns out that 
these can be substantial, since $W_1$ is of dimension $4$ and
therefore of the same order as the non-derivative operator $W' =
(H_1 H_2)^2/M_* $ considered in \cite{strumia,dst}. To investigate
these effects, first notice that despite the presence of the derivative
in the superpotential, the auxiliary fields of $H_1, H_2$ 
are not dynamical and can be eliminated by their eqs of motion.
After doing so, one finds a Lagrangian for the scalar components
\medskip
\bea
\cL= -h_i^\dagger \,(\Box +\xi_3
\Box^2-2\,\mu\,\sqrt\xi_3\,\Box)\,h_i-V,\qquad i=1,2
\eea
where $V$ is that of the MSSM. Finding the extrema of the
potential, going to the  mass eigenstates $(h,H)$ etc, proceeds as in the MSSM, while
the kinetic terms are invariant under these transformations. One then
finds  the poles of the propagators above from
\bea
-p^2+\xi_3 p^4+2\,\mu\,\sqrt\xi_3 \,p^2 + m_h^2=0,
\eea
where $m_h$ is the value computed in the MSSM. With $p^2=m_h^2+\delta
m_h^2$, the effect of $W_1$ on the lightest Higgs mass is found to be
 \bea 
\frac{\delta  m_h^2}{ m_h^2} \approx 2\,\mu\,\sqrt\xi_3 \,
=\frac{2\mu}{M_*}. \ 
\label{mssm5} 
\eea 
This correction is  of order $10\%$  for $M_*\sim 10$ TeV and $\mu\sim
500$ GeV and can 
therefore increase the Higgs mass above the
experimental limit even before adding the quantum corrections~!
Such a correction is comparable to the one found  in \cite{strumia,dst}
using the operator $(H_1 H_2)^2/M_*$.
%%% for large $\tan\beta$ region.
For a further discussion  of these corrections see also \cite{ADGT}.

\section{Conclusions}

Higher dimensional  operators (derivative or otherwise)
are a common presence in 4D effective, nonrenormalisable theories.
They are easily generated in  the low-energy effective action
from 4D renormalisable theories upon integration out 
massive (super)fields.
Such operators are also  dynamically generated by radiative corrections
in effective theories of compactification even for the
simplest orbifolds. 

Motivated by this, we investigated in
detail  the case of 4D N=1 supersymmetric models with such operators.
Using a superfield language it was shown that  a 4D N=1 supersymmetric
theory with higher derivative terms in the Kahler potential and an arbitrary
superpotential is equivalent to a 4D N=1 theory of second order with
two  additional superfields and renormalised  interactions.
Because in the initial, higher derivative formulation  of the theory
both $\Box \phi$ and the
auxiliary field $F$ where propagating,  by supersymmetry this lead 
in the two-derivative formulation of the theory to the existence
 of the two additional superfields mentioned above.

The method developed was then
extended to the case of 4D N=1 models with higher derivative
terms in the superpotential whose remaining part is otherwise
arbitrary. It was again showed that such model is equivalent
to a 4D N=1 second order theory with an additional (ghost)
superfield and renormalised couplings. Unlike the case of higher
derivatives in the Kahler potential, in this case there is only one
additional superfield in the second-order formulation because in this
case  only $\Box\phi$ is propagating in the higher derivative theory,
and this implied, by supersymmetry, the existence of one additional
(ghost) superfield (indeed, we found that 
$\tilde\Phi_3$ was  acting only as a constraint superfield
in Section~\ref{sec3}, whereas it was a propagating degree of
 freedom in Section~\ref{sec1}).
Finally, it was verified in both cases 
that in the second order formulations of the theory
the spectrum must be computed with  an appropriate 
metric in field space to  account for the  negative
 kinetic terms of the ghosts fields.

In both cases the  couplings of the new, second order theory,
acquire already at the tree level a dependence on the scale of
the higher dimensional operator.
 The new, second order formulation of the theory has the advantage
 of providing  a  standard, familiar
 framework for investigating the role of these
 operators in explicit models. We argued that
if there are no additional operators of $D>4$ in the original theory
apart from the higher derivative ones considered,
 the second order formulation 
of the theory has only D=4 operators. This theory can be
renormalisable,  under some assumptions for
the analytical continuation from the Minkowski to Euclidean metric.
This requires that the ghost propagators be Wick rotated
to Euclidean space in the same sense as the particle-like ones, 
leading to similar UV for both  Minkowski and Euclidean descriptions.

In the new basis of the second order theory,
the original superfield is a mixing of particle and ghost-like
superfields, and thus the particle-like degrees of freedom
are not identical in the two descriptions.
This  brings an intriguing issue, regarding which of the two
descriptions is more fundamental. Ultimately this refers to which
 choice of the degrees of freedom one should make for the particle-like
field: original field of the fourth order theory
or the particle-like degree of freedom in the second order theory.
This issue is relevant particularly at the loop level,
when superfields re-scaling anomalies associated with
the transformations we performed, can bring in
quantum corrections to the equivalence of the two formulations.

Our analysis remains valid in the presence of supersymmetry breaking
terms, as it was confirmed by computing the spectrum in both formulations
of the theory with higher derivative operators, for explicit forms
of soft breaking terms. The higher derivative operators are also
important for supersymmetry breaking. We showed that models with
higher derivative terms which look rather un-interesting in the
original formulation and could be disregarded when decoupling these
terms, turn out to be in the two-derivative formulation,
interacting theories that exhibit spontaneous supersymmetry breaking
\`a la O'Raifeartaigh.

The analysis can be applied in the presence of arbitrary higher
derivative terms, using eventually an iteration of the method
presented in Sections~\ref{sec1} and \ref{sec3}.
 Higher dimensional Kahler terms other that those 
leading to standard kinetic terms in the two-derivative formulation,
can be ``moved'' into the superpotential with an additional
(super)derivative and become, in the new field basis, higher
dimensional non-derivative interactions. An example of this type 
was discussed in the second part of Section~\ref{SSB}.
Similar techniques can be applied in the case of complicated 
derivative interactions in the superpotential, by replacing
derivatives  of  superfields with new superfields and appropriate 
holomorphic constraints in the Lagrangian. Finally, in specific
cases and for appropriate parameters in the original theory, 
some of the  Kahler terms of  ghost superfields that can emerge
in the two-derivative formulation   may in some cases decouple from 
the Lagrangian, if these fields do not have  superpotential terms. 
To conclude, our method shows that  theories with higher  derivative 
operators can be mapped to theories with  higher dimensional,
non-derivative  operators.

An application to the case of higher derivative terms in the 
MSSM Lagrangian was also  presented. It was estimated  that 
the Higgs mass can be lifted above the experimental limit by such 
terms, even before adding quantum corrections associated with them.
Regardless of the exact nature of the ghost fields 
that higher derivative operators bring in (i.e. whether these
fields exist as asymptotic states or only loop ones),  
the method  we presented can be used as a perturbative tool 
to investigate the effects on low energy physics of new high-scale physics
associated with  higher  derivative operators, 
much in the same way this is done for higher dimensional operators.

In general the presence of ghost superfields in a 4D N=1 action
has as effect that  the scalar potential of such theory is not positive
definite. Therefore, the  vanishing of $V$ is not equivalent to exact
supersymmetry anymore and one
can have $V>0$, $V<0$ and even $V=0$ for broken supersymmetry.
In the last case $\vert \tilde F_i\vert=\vert \tilde F_j\vert\not=0$ where
$i$ and $j$ label the contributions of particle and ghost-like
states to $V$. A vanishing scalar potential would require the
breaking of supersymmetry be done by both the ghost and particle-like degrees of
freedom.  This last case could be of some interest for
the cosmological constant problem.

We would like to end our  discussion of the equivalence on the
two formulations of the theory with higher dimensional operators
 with the following observation.
The equivalence we showed between the fourth order and second order
formulation is valid at the classical level.
A legitimate question is then  whether one can make similar claims
of equivalence at the quantum level.  
Although the question is beyond the purpose of this paper,  let us make the
following remark.
The study of the quantum equivalence is somewhat beyond the possibility of
 an {\it effective} field theory framework, where the absence
of a detailed UV completion would render such analysis 
incomplete or valid in very specific cases only.
 Nevertheless, restricting ourselves to the 
lagrangian with one higher derivative operator, 
we performed an explicit check of the equivalence at 
the one-loop level, for the radiative correction to the
mass of the original scalar field $\phi$, after a soft 
breaking of supersymmetry. Using (\ref{originalL}) and its second order 
formulation  (\ref{genW}) with (\ref{phiphid}), (\ref{mu2}),
we checked explicitly  that one obtains the same one-loop result. This is 
interesting in itself and checks the validity of our formalism at
the quantum level too, for this particular case. However, given the
effective character of these theories and the absence of a UV
completion, one should be
careful not to extrapolate this finding to a  general, similar
statement of quantum equivalence of the two formulations.

\section*{\bf Acknowledgements}
This work was partially supported by INTAS grant, 03-51-6346,
RTN contracts MRTN-CT-2004-005104 and MRTN-CT-2004-503369, CNRS PICS
\#~2530,  3059 and 3747, and by a European Union Excellence Grant,
MEXT-CT-2003-509661. D.G. also acknowledges the partial
support of the Marie Curie Research Training Network
of the European Community, contract n. MRTN-CT-2006-035863
and that of a CERN visiting fellowship.

\section*{Appendix}

In the case of higher derivatives of Section \ref{sec1}
the eigenvalues were
\bea
\nu_1=\frac{1}{2}\,  (1+\sqrt\eta    ),\qquad
\nu_2=\frac{1}{2}\,  (1-\sqrt\eta    ),\qquad
\nu_3=-\frac{\xi \,m^2}{16},\qquad(\eta\equiv 1+4\,\xi\,m^2)
\eea
The corresponding  eigenvectors $v_{ij}$ with $\Phi_i'=v_{ij}\,\Phi_j$
(see Section~\ref{hdosec})
are respectively
\medskip
\bea
v_{1j}&=&\frac{1}{\vert\vert v_1\vert\vert}\,\Big\{
- e^{i\,h_1}\,\frac{\nu_1}{m\,\sqrt\xi} \cos\theta,
- e^{-i\,(h-h_1)}\,\frac{\nu_1}{m\,\sqrt\xi} \sin\theta,
1\Big\}_j \,
\nonumber\\[12pt]
v_{2j}&=&\frac{1}{\vert\vert v_2\vert\vert}\,
\Big\{
-e^{i\,h_1}\,\frac{\nu_2}{m\,\sqrt\xi} \cos\theta,
- e^{-i\,(h-h_1)}\,\frac{\nu_2}{m\,\sqrt\xi} \sin\theta, 1\Big\}_j\,
\nonumber\\[12pt]
v_{3j}&=& \frac{1}{\vert\vert v_3\vert\vert}\,
\Big\{-e^{i\,h}\,\tan \theta,1,0\Big\}_j,\qquad j=1,2,3.
\eea

\medskip
\noindent
with $\vert\vert v_i\vert\vert$, $\,i=1,2,3$, the norm
 of the corresponding vector.

In the case of higher derivatives of Section~\ref{sec3}
the eigenvalues were
\medskip
\bea
\nu_1=\frac{1}{2}\,  (1+\sqrt{\eta^\prime}    ),\qquad
\nu_2=\frac{1}{2}\,  (1-\sqrt{\eta^\prime}    ),\qquad
\nu_3=0,\qquad\eta^\prime\equiv 1+4\,\xi\,m^2 (1+ s_2^2/16)
\eea

\medskip
\noindent
The corresponding  eigenvectors used there were

\bea\label{spthdo}
v_{1j}&=&\frac{1}{\vert\vert v_1\vert\vert}\,\Big\{
\frac{- \nu_1\,e^{i\,h_1}}{m\sqrt\xi}\,\cos\theta+\frac{e^{-i\,(h_1-h)}
\,s_2}{4}\,\sin\theta,\,
\frac{-s_2\,e^{-i\,h_1}}{4}\cos\theta
-\frac{\nu_1\,e^{-i\,(  h -  h_1)}}{m\sqrt\xi}\sin\theta,\,1
\Big\}_j
\nonumber\\[12pt]
v_{2j}&=&\frac{1}{\vert\vert v_2\vert\vert}\,
\Big\{
\frac{-\nu_2\,e^{i\,h_1}}{m\sqrt\xi}\cos\theta
+\frac{s_2\, e^{-i\,(h_1-h)}}{4\,}\sin\theta,\,
\frac{-s_2\,e^{-i\,h_1}}{4\,}\cos\theta-\frac{\nu_2\,e^{-i\,(h-h_1)}}
{ m\,\sqrt\xi} \sin\theta,\,1\Big\}_j
\nonumber\\[12pt]
v_{3j}&=& \frac{1}{\vert\vert v_3\vert\vert}\,
\Big\{
\frac{-4}{s_2}\,e^{i\,(h-h_1)}\,\sin\theta,\,\frac{4}{s_2}\,e^{-i\,h_1}\,
\cos\theta,\,1\Big\}_j
\eea

\end{document}